\documentclass{article}

\usepackage{mathtools}
\usepackage{array}
\usepackage{float}
\usepackage{easy-todo}
\usepackage[none]{hyphenat}
\usepackage{flafter}
\usepackage{placeins}
\usepackage{color}
\usepackage{times}
\usepackage{cite}
\usepackage{amsmath}
\usepackage{latexsym}
\usepackage{pstricks}
\usepackage{setspace}
\usepackage{subfig}

\usepackage{fancyhdr}
\usepackage{array}
\usepackage{float}
\usepackage{setspace}

\usepackage{graphicx}

\usepackage{acronym}
\usepackage{url}
\usepackage{easy-todo}
\usepackage{subfig}
\usepackage{multirow}
\usepackage[left=1.25in,right=1.25in,top=1in,bottom=1in]{geometry} 
\usepackage{algorithm}
\usepackage{algorithmic}
\pdfoutput=1
\title{GSM based CommSense system to measure and estimate environmental changes}

\author{\textbf{Abhishek Bhatta}\\
Electrical Engineering Department\\
University of Cape Town
\\
Email: bhattaomatic@gmail.com\\
\and
\textbf{Amit Kumar Mishra}\\
Electrical Engineering Department\\
University of Cape Town
\\
Email: akmishra@ieee.org}

\begin{document}

\maketitle
\begin{abstract}
Facilitating the coexistence of radar systems with communication systems has been a major area of research in radar engineering. 
The current work presents a new way to sense the environment using the channel equalization block of existing communication systems. 
We have named this system CommSense. 
In the current paper we demonstrate the feasibility of the system using  Global System for Mobile Communications (GSM) signals. The implementation has been done using open-source
 Software Defined Radio (SDR) environment. 
In the preliminary results obtained in our work we show that it is possible to distinguish environmental changes using the proposed system. 
The major advantage of the system is that it is inexpensive as channel estimation is an inherent block in any communication system and hence the added cost to make it work as an environment sensor is minimal. 
The major challenge, on which we are continuing our work, is how to characterize the features in the environmental changes. This is an acute challenge given the fact that the bandwidth available is narrow and the system is inherently a forward looking radar. 
However the initial results, as shown in this paper, are encouraging and we intend to use an application specific instrumentation (ASIN) scheme to distinguish the environmental changes. 
\end{abstract}

\section{Introduction}\label{sec:intro}
 Classical radar systems have been designed primarily for military operations. 
Of late many interesting ways of using radio-frequency spectrum for radar purpose have been under research. 
One such  concept is commensal\footnote{The word commensal has been borrowed from biology in which this represents co-existence of two species out of which one is benefited and the other remains unaffected.}  or passive radar, which uses signals of opportunity to detect targets without affecting the functionality of the parent system
\cite{inggs2013commensal,inggs2012modelling,inggs2014planning,tan2005passive,tan2003feasibility,berger2010signal,sun2008aircraft}. 

One implementation of commensal system is the recently active area of Passive Bistatic Radar (PBR) \cite{griffiths2009passive,krysik2011ground,samczynski2011concept,coleman2008passive,homer2002passive}.
 
 In our work we shall describe a kind of commensal radar which we call communication sensor (CommSense) \cite{mishra2015commsense,mishra2015patent} system. 
It uses the channel estimation processing in communication systems to estimate changes in the environment. 
This  system can, potentially, be used to monitor land terrain, sea state or even natural disasters. 
 The novelties of the proposed system are as follows.
\\
 First of all, this system is built upon existing communication systems. The cost of implementing such a system is therefore considerably low. 
Secondly, unlike PBR systems it does not process the information using correlation, rather it uses a known training sequence to estimate changes in the environment. It therefore eradicates the need for a reference antenna pointing towards the transmitter. We have implemented the prototype system using open source GNURadio software and Nuand BladeRF$\times 40$ SDR hardware\cite{url:gnuradio,url:bladeRF}.

 It must, however, be noted that there are two major challenges in implementing this  system. 
 First of all, the conventional concepts of resolution will not be valid here mainly because of the fact that the bandwidth available to us is very narrow. 
This makes the measurement system a severely ill-posed inverse problem. 
Secondly, this is a forward looking and non-coherent radar system which theoretically limits the amount of information it can capture. 
 We hypothesize that these two problems can be solved using the application specific instrumentation framework \cite{mishra_10_asin, sardar_14}.
This will be our future work, whereas in this paper we demonstrate the feasibility of the concept of sensing the environment from communication signals.
 
 The rest of the paper is organized as follows.
 Section \ref{sec:sysdes} gives a basic understanding of GSM protocol and an introduction to the design of this system. Section \ref{sec:realimple} provides details about the real time implementation of the system. Section \ref{sec:data} shows how the data analysis is done, specifically focussing on the Probability Distribution Function (PDF) analysis and clustering. Finally Section \ref{sec:conc} concludes the work and shows the scope for future developments.

\section{GSM Channel Estimation and CommSense System Design}\label{sec:sysdes}

 Any signal transmitted over the channel gets affected by the channel itself. 
This effect is minimized in wireless communication systems using post-reception processing blocks called channel estimation and equalization \cite{pukkila2000channel,pu2010channel}. 
 The hypothesis of our work is that the channel information from the received signal can be harnessed and used to statistically monitor the changes in the environment. For example we will be able to differentiate between different weather conditions or produce an alternate model (which, potentially, can give more information) for sea waves and mountainous terrains. 
 We call this system CommSense system. 
The reader is suggested to refer \cite{mishra2015commsense} for system level information of a generic CommSense system. 

 The communication technology infrastructure used for the current implementation of the CommSense is GSM. 
 This is because it is currently the most widely used wireless communication technology in South Africa. 
We are also working towards investigating its feasibility using LTE infrastructure \cite{sardar_15_lte}. 
 GSM provides coverage of almost the entire nation. It is a wireless communication protocol developed by European Telecommunications Standard Institute (ETSI) to provide a standard in the wireless telecommunication industry. 
CommSense system relies on the broadcast signals transmitted by the base stations for phones to detect and connect to the particular base station, thus not hampering the communication system in any way. 
GSM transmits a known training sequence in every frame so that the receiver is able to detect and cancel the channel interference. 
Using the information from this, the channel states can be estimated. 

\begin{figure*}[t]
  \begin{center}
  \includegraphics[width=0.9\textwidth]{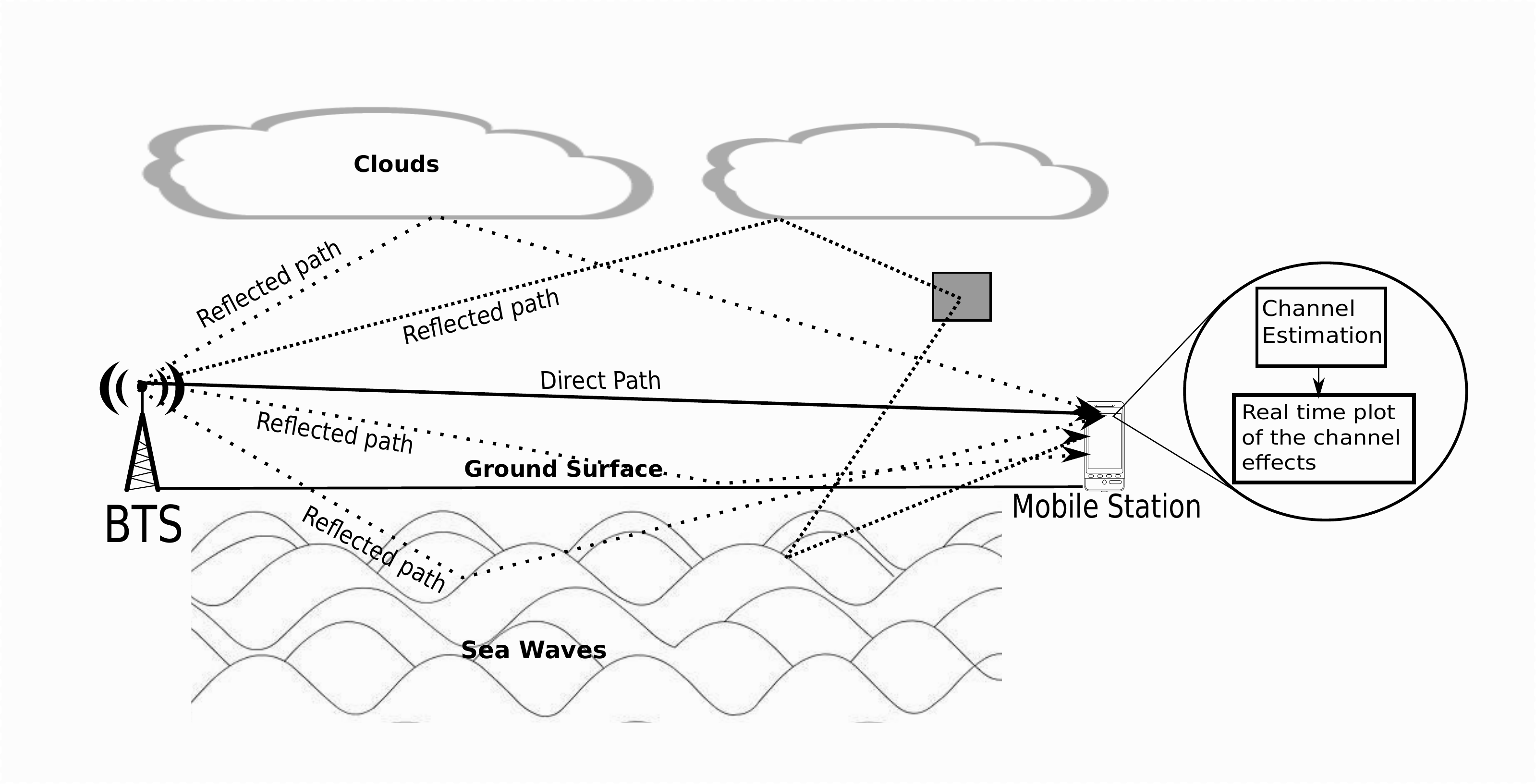}
  \end{center}
  \caption{System Overview.}\label{fig:1}
\end{figure*}

 Figure \ref{fig:1} shows the concept diagram of the application. 
Signal from the base station reaches the mobile station through multiple different paths and is affected by the physical properties of each path. 
Out of all these returns, the signal with highest Signal to Noise Ratio (SNR) is extracted by the method of preliminary channel equalization and used for the purpose of communication. In our work we shall be using this preliminary channel equalization block. 
It, however, can be noted that there are further blocks in the processing chain in the receiver which use a range of algorithms to take care of channel noise. For the purpose of our work we shall use {\it channel equalization} to mean this {\it preliminary channel equalization} processing. 
In order to implement channel equalization the communication system transmits a known sequence of bits with every frame. These known bits, also known as training sequence, are extracted from the channel convoluted received signal and used to extract The difference between the known and expected signal. This difference between the known and the received signal gives the state of the channel through which the signal has travelled. In the proposed system this estimated information is used to find the changes in the channel state in different environmental conditions.

\subsection{System Design}

 CommSense system acts as a commensal radar, eliminating the need of a new transmitter. As the processing power of smart-phones has significantly improved, it is planned to eventually implement CommSense as a mobile application. 
 For the purposes of proving the hypothesis it has been implemented on the SDR platform with open source software. The hardware used for the implementation is BladeRF$\times 40$ manufactured by Nuand \cite{url:bladeRF}. 
 It is a Universal Serial Bus (USB) powered SDR board with an on-board transmitter and receiver port. It can operate over a frequency range of 300 MHz to 3.8 GHz which gives the opportunity to implement the proposed system using GSM technology. The software used for implementing the real time channel estimation system is GNU Radio \cite{url:gnuradio}. It is an open source software tool-kit, distributed under GNU general public license, that provides the opportunity to implement signal processing blocks using open source RF hardware for research purposes. 
 
 GSM is a wireless communication protocol that acts as a host to this system \cite{digitalcelltel,radiotransrec,hillebrand2002gsm}. It uses Frequency Division Multiple Access (FDMA) and Time Division Multiple Access (TDMA) in order to accommodate multiple users. 
 The various frequency bands of operation for this technology are 800 MHz, 900 MHz, 1800 MHz, 1900 MHz. 
 GSM works in frequency duplex mode thus having a different frequency for uplink and downlink. 
 GSM900, used here for data collection, uses 880-915 MHz for uplink and 925-960 MHz for downlink communications. These bands are further divided into 200 kHz bands and are separated by a number called Absolute Radio Frequency Channel Number (ARFCN). 
 Each wireless channel is divided into TDMA frames with 8 time slots of 577 $\mu $s each. Each timeslot is time shared between the mobile and the base station. The logical channels are piggybacked on the physical channel, described in \cite{digitalcelltel}.
  
 Each frame in the GSM system transmits a known training sequence for the receiver to detect and reduce the possibility of transmission error. 
 This functionality provided by GSM is used in this project to estimate the channel effects on the signal and analyse it to find the differences in the various types of terrain and environmental conditions. 
 A frame in GSM system is shown in Figure \ref{fig:2}.
 This structure shows the normal burst that has been used for the extraction of the parameters. 
 This burst has 114 bits of information transmitted with tail bits, flags and a guard period to identify the start and stop of the burst and different datasets. The training sequence is 26 bits long out of which the central 16 bits are used for channel equalization. 
 In the current implementation these equalized channel values are used to determine the surrounding environment. A guard period is added at the end of each frame to provide a window of error against distortions that occur due to the rise and fall time of the signal.

\begin{figure}[t]
  \begin{center}
  \includegraphics[width=0.9\textwidth]{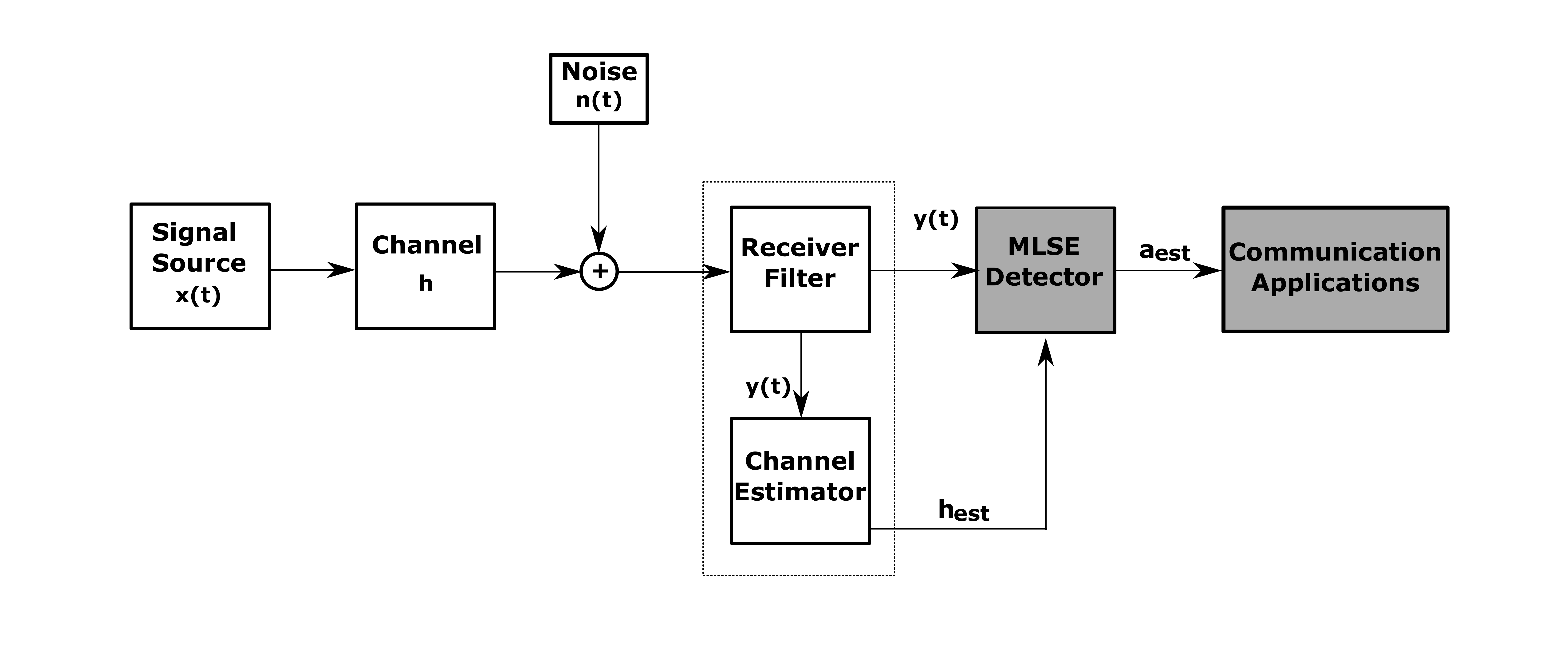}
  \end{center}
  \caption{Channel Estimation Block Diagram.}\label{fig:3}
\end{figure}

\subsection{Channel Equalization Filter}
\begin{figure}[b]
  \begin{center}
  \includegraphics[width=0.9\textwidth]{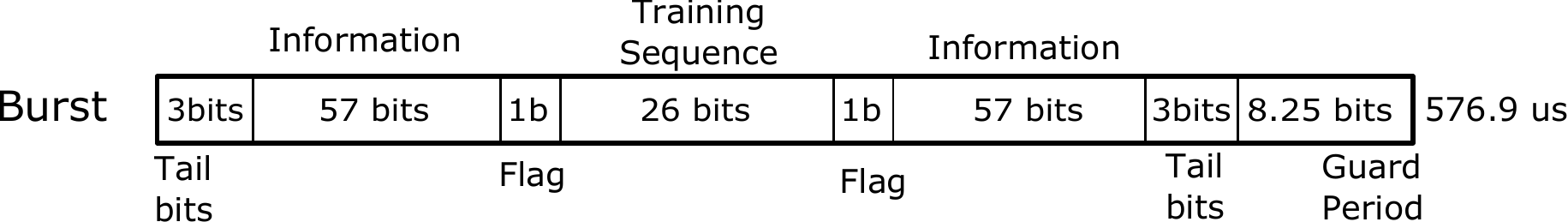}
  \end{center}
  \caption{GSM Normal Channel Frame Structure.}\label{fig:2}
\end{figure}

 Figure \ref{fig:3} shows the block diagram of the implementation. 
 The received GSM signal can be mathematically represented as equation (\ref{eq:1}). Here, the received signal $y(t)$ is the convolution of the transmitted signal $x(t)$ and Channel Impulse Response\footnote{CIR is a black-box linear model of the channel effects.} (CIR) $h(t)$ in the presence of Additive white Gaussian noise (AWGN) $n(t)$. 
 The transmitter sends a known training sequence in each frame as shown in Figure \ref{fig:2}, which is divided into reference length of $P$ and guard period of $L$ bits \cite{pukkila2000channel}. This equation can be represented in the matrix form and is shown in equation (\ref{eq:2}). 
\begin{align}
y(t) &= h(t)*x(t) + n(t)\label{eq:1}\\
\textbf{y} &= \textbf{Mh} + \textbf{n}\label{eq:2}
\end{align}

 In equation (\ref{eq:2}) \textbf{M} represents a Toeplitz 
like matrix structure of the given training sequence as shown in equation (\ref{eq:4}). 
This matrix is made of the \textbf{m} array that is shown in equation (\ref{eq:3}), which is actually the oversampled information from each received frame after filtering out the Gaussian 
noise.

\begin{equation}\label{eq:3}
\textbf{m} = [m_0\quad m_1\quad ....\quad m_{P+L-1}]^T
\end{equation}

\begin{equation}\label{eq:4}
\textbf{M}_{P+Cl-1 \times Cl}= \begin{bmatrix}
m_0 & 0 & 0 & \cdots & 0\\
m_1 & m_0 & 0 & \cdots & 0\\
m_2 & m_1 & m_0 & \cdots & 0\\
\vdots & \vdots & \vdots & \vdots & \vdots\\
m_P & m_{P-1} & m_{P-2} & \cdots & m_{0}\\
0 & m_{P} & m_{P-1} & \cdots & m_{1}\\
0 & 0 & m_{P} & \cdots & m_{2}\\
\vdots & \vdots & \vdots & \vdots & \vdots\\
0 & 0 & 0 & 0 & m_{P}\\
\end{bmatrix}
\end{equation}

\begin{equation}\label{eq:5}
h = [ h_0 \quad h_1 \quad h_2 \quad \cdots \quad h_n]^T
\end{equation}

 In equation (\ref{eq:4}), the matrix \textbf{M} is a $P+Cl-1 \times Cl$ matrix where $P$ is the reference length of training sequence and $Cl$ is the length of the CIR. equation (\ref{eq:5}) shows the CIR array, here each value of the array represents the channel parameters of one reflected signal. Thus the length of the array determines how many multi-path signal are being analysed. The Least Squares algorithm finds the CIR by minimising the squared error quantity which in the presence of white Gaussian noise gives equation (\ref{eq:6}). 
\begin{equation}\label{eq:6}
\textbf{h}_{est}= (\textbf{M}^H\textbf{M})^{-1}\textbf{M}^H\textbf{y}
\end{equation}
Here, $\textbf{M}^H$ denotes Hermitian transpose matrix and $()^{-1}$ denotes matrix inverse. In equation (\ref{eq:6}) we can observe that 
the received signal matrix \textbf{y} is multiplied to a matrix also known as the pseudo inverse matrix of \textbf{M}.

\begin{figure}[h]
  \begin{center}
  \includegraphics[width=0.8\textwidth]{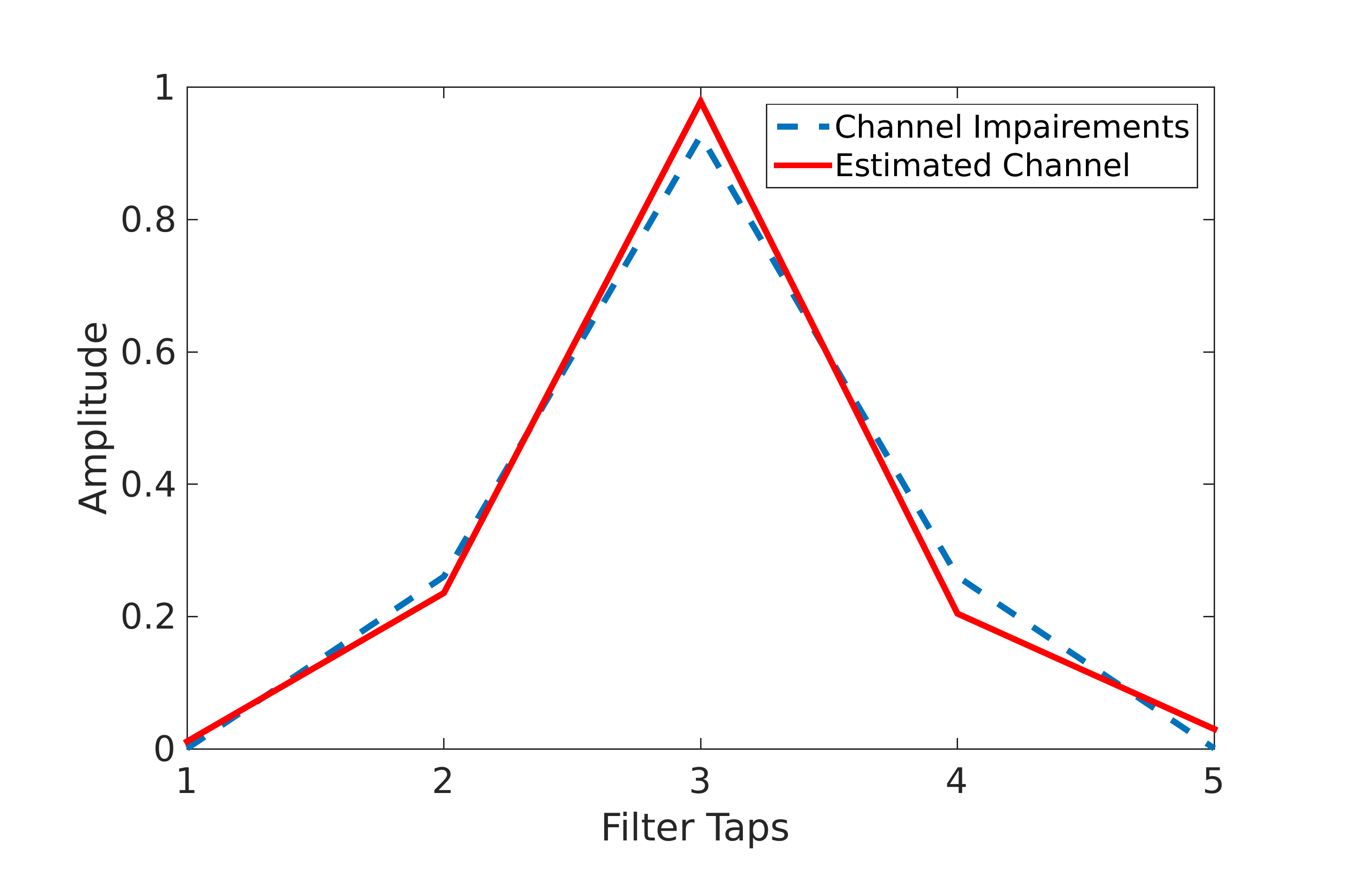}
  \end{center}
  \caption{Simulated Least Square Channel Estimation in presence of AWGN.}\label{fig:4}
\end{figure}

\begin{figure}[t] 
	\centering
	\includegraphics[width=0.9\textwidth]{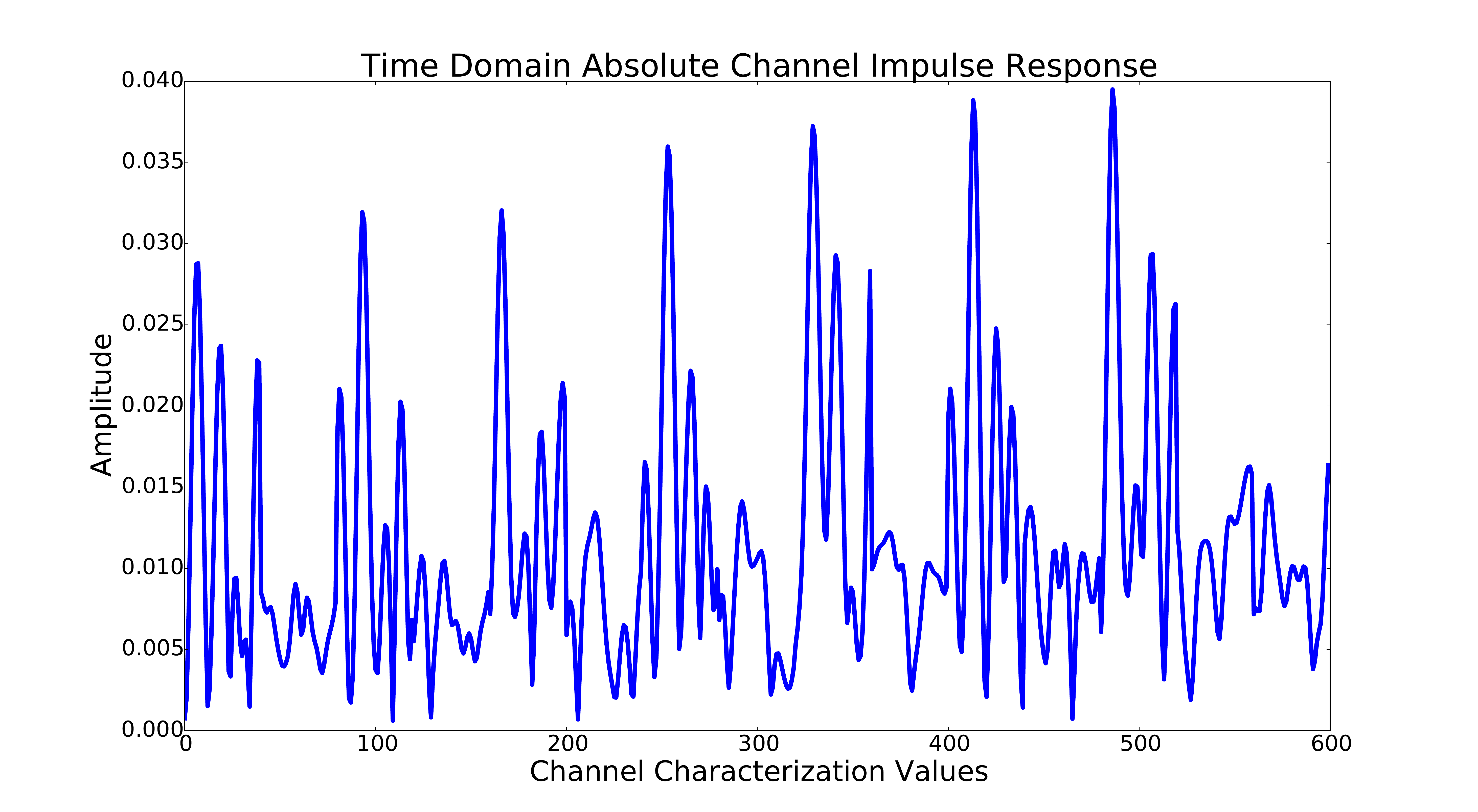}
	\caption{Estimated Normal channel absolute values in terms of the $I$ and $Q$ component, in this figure is represented by  $h_{est}=\sqrt{I^2+Q^2}$.}\label{fig:5a}
\end{figure}

 Figure \ref{fig:4} shows the plot for simulated CIR in the presence of AWGN. This result is for a single GSM frame being transmitted over a known channel of length 5 taps. The estimated value is very similar to the original CIR value which shows that the algorithm is working properly. Then this is implemented in real-time as shown in Figure \ref{fig:5a}.

\section{Real Time Implementation} \label{sec:realimple}

 The system is implemented in GNU Radio, which does the channel estimation and plots the CIR in real time as shown in Figure \ref{fig:5b}. 
 The hardware used is BladeRF$\times 40$, which works at a frequency band of 300 MHz to 3.8 GHz with a maximum physical bandwidth of 28 MHz. The antenna used for this specific implementation is an off the shelf quad band GSM antenna. The accuracy of the system is proved by decoding the received bit-stream and analysing it in Wireshark (a packet analyser tool), which gives the basic information as the location of the base station, carrier service provider and so on.
 
\subsection{Extracting the Channel Values}
 The steps taken to implement the channel equalizer in real-time are mentioned here. Correlation between the received signal and the known training sequence is used to extract the differences between the two signals and this information is saved in a file for further analysis.
 
 The SDR hardware, bladeRF, receives the GSM signal downconverts it according to its Phase Locked Loop (PLL) clock frequency and gives out the In-phase ($I$) and the Quadrature ($Q$) components of the signal. The receiver first needs to get synchronized with the base station, the Synchronization Channel (SCH) burst is used for this purpose. The SCH burst consists of a 64 bit long training sequence and transmits the same sequence in every timeslot allocated to it, thus it is easily distinguishable and the receiver can get synchronized. Once the receiver synchronizes with the GSM base station, the frequency offset is calculated using the Frequency Correction channel (FCCH). 

 The received burst type is identified in two steps first with the help of the carrier index of the burst and then with the burst number. Carrier index zero (C0) is a special case where only Broadcast Channel (BCCH) data is transmitted. BCCH carries a repeating pattern of system information such as identity, configuration and available features for the base station. In case of other carrier indices the burst type is determined by the burst number. The most commonly transmitted burst is the normal burst thus in this implementation we are focussing on extracting the channel information from the normal burst.

 In order to receive a burst completely the receiver needs to wait for two consecutive guard periods. Although there will be an overlap between the guard periods of two consecutive bursts, this is necessary to ensure full burst is received. Once a normal burst is identified the start and the stop position of the burst is determined eliminating the guard period as it consists of 8.25 bits on each side and is distinguishable. The central 26 bits of the remaining bits are extracted as it constitutes of the received training sequence. The training sequence extracted here is a complex data so the already available sequence also needs to be converted into a complex number. The two complex training sequences are correlated with each other and the correlation information is saved in a buffer. A pointer is placed at the beginning of the buffer and based on predefined CIR length the pointer is moved from the beginning to the length of the CIR. The values corresponding from the beginning to the CIR length is saved in an array and then moved to a file where it is saved for further analysis. In case of communication systems the maximum absolute value of the CIR is chosen discarding the rest of the values and that information is used to equalize the channel and get better communication. For the purpose of this implementation the entire channel information is necessary thus all of it is saved. 
 
\begin{figure}
	\centering
	\includegraphics[width=\textwidth]{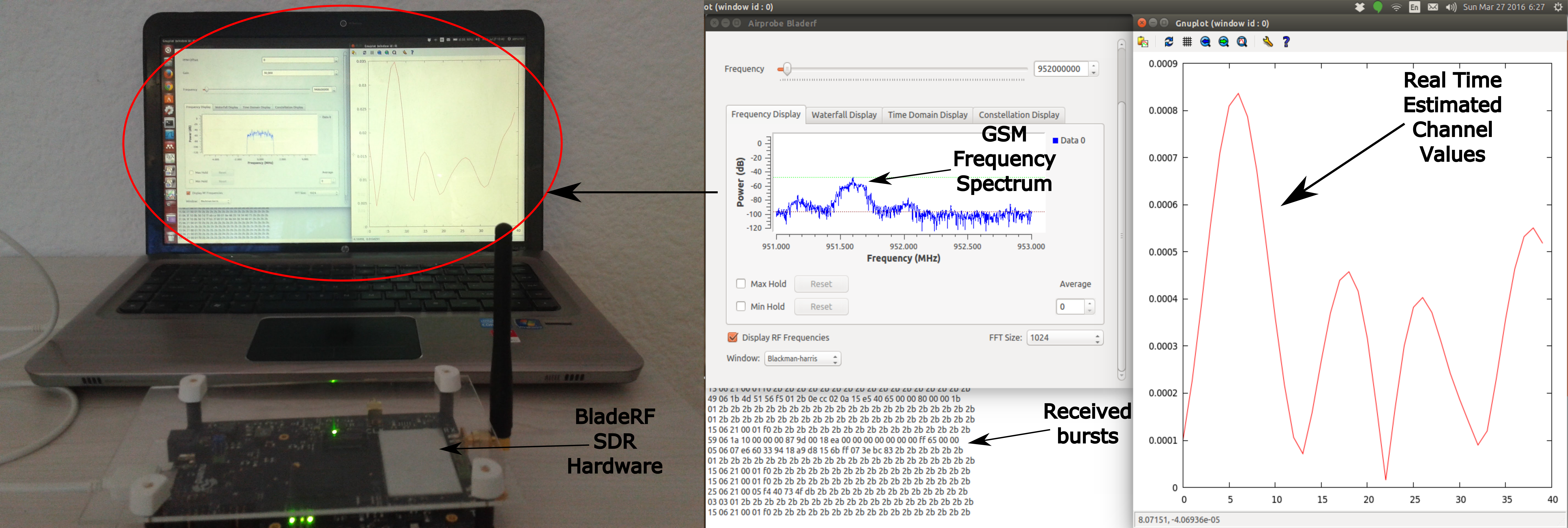}
    \caption{GNU Radio implementation of Channel Equaliaztion with BladeRF$\times 40$.}\label{fig:5b}
\end{figure}

 In order to check if the receiver is working as intended the entire receive structure is made referring to the gr-gsm libraries available for GNU Radio and the bursts are printed out using a message printer. Figure~\ref{fig:5b} shows the plot of the entire implementation including the GSM frequency spectrum, received bursts and the real time estimated channel values. 
 
\subsection{Real Time Channel Estimation and Data Capturing}

 Figure \ref{fig:3} shows the entire flow of the implementation, where the signal source is the transmitting base station and the processing is done on \textit{over the air} signals. The receiver filter block removes most of the noise from the signal and shows the GSM frequency spectrum as shown in Figure \ref{fig:5b}. The data is used to estimate the channel and store the values in a file for further analysis. The darker blocks in Figure \ref{fig:3} are the normal GSM processing blocks which are not affected by this system as it runs parallel to the communication system.
 
 One of the major challenges faced during the implementation is finding the start-point of the burst. This is done with the help of the guard period. 
 The receiver waits for two consecutive guard periods, once received it considers the start point of the burst. In the case of an error in receiving the guard periods, the system drops the current set of data and waits for the next set and then repeats until it gets two consecutive guard periods, thereby eliminating the chances of any ambiguous data.

\section{Analysis of Captured Data}\label{sec:data}

 To prove the hypothesis the real-time channel information is captured in different locations and classified as Statistical Characterization, Distinguishing Events and Special case. 
 Table \ref{tab:captures} lists details of all the datasets collected and analysed in this work. 
 The estimated channel information is analysed to show the changes in the channel state in different environmental condition. 
Three different types of analysis are presented in this section, two in statistical domain and one in data/time domain. 
 In the first approach we estimate the PDF of the channel values captured at different scenarios. This shows that the estimated PDF's vary as the environmental conditions change which means that a ``hypothesis test'' kind algorithm can be used to distinguish different environment types from the estimated channel values. 
 In rest of the paper we shall use simply PDF to mean estimated PDF.
In the second approach we perform Chi-square test to assess the goodness of fit between the observed data and the theoretically expected dataset.
 In the third approach we check for the clustering of the estimated channel values for different capture scenarios. 
 For easier visualisation we used principal component analysis (PCA) to reduce the dimension of the data. This shows that the data are highly clustered in PC domain. 

\begin{table}[tbph]
\resizebox{0.95\textwidth}{!}{\begin{minipage}{1.1\textwidth}
\centering
\begin{tabular}{ | c | p{35mm} | p{80mm} | } \hline
\multicolumn{3}{|c|}{\textbf{Captured Data}} \\ \hline
\textbf{Type} & \textbf{Location}  & \textbf{Comments}\\ \hline
\multirow{5}{*}{\textbf{Statistical Characterization}} & Sea Beach & To check the effects of the sea waves on the signal, multiple locations may also be tested \\ \cline{2-3}
 & Hill next to a road & Check the effects on the signal in the presence of a hill and see repeatability (safe location still needs to be decided) \\ \cline{2-3}
 & Highland & Captures taken at the road next to a steep terrain \\ \cline{2-3} 
 & Heavy Rain, slightly humid climate, hot day & Captures taken at the same location at different instances of the day showing the differences due to different weather conditions\\ \cline{2-3}
 & Parking space & Captures are taken on different times of the day. One set showing when the place is full and the other showing empty parking lot\\ \hline
\multirow{3}{*}{\textbf{Distinguishing events}} & Train Station & At a train station with and without a train in proximity\\ \cline{2-3}
& J stairs & At the stairs to the entrance of an approximately 10 meter tall building.\\ \cline{2-3}
& Bus & With and without the presence of a bus in the proximity, two different sets of captures were taken at different locations\\ \hline
\multirow{2}{*}{\textbf{Special case}} & Car & Captures taken with and without the presence of a car in proximity of the receiver \\ \cline{2-3}
 & Corner Reflector & Captures are taken when a corner reflector is placed at a distance of 2 meter from the antenna in various different configurations 'V' is vertical, 'H' is horizontal, 'VH' refers to a dihedral corner reflector\\ \hline
\end{tabular}
\end{minipage}}
	\caption{The different sets of captured data.}\label{tab:captures}
\end{table}

\subsection{PDF Analysis}
In this set of analysis we plot the empirical PDF of the data collected from different scenarios as described in Table~\ref{tab:captures}. 
In addition to the empirical PDF we also plot the PDF of the data for four different theoretical PDF models, i.e. Rayleigh, Normal, Log-normal and Gamma distributions. 
A brief description of these distributions and maximum likelihood estimation (MLE) expression of their parameters are given in Appendix I. 

 Figure \ref{fig:hist_rep1} and \ref{fig:hist_rep} show the distribution plot of two different datasets in similar conditions. Both the captures are taken near the entrance of a building about 10 meters tall. The difference between the two captures is the wind speed. During the capture of Figure \ref{fig:hist_rep1} the wind speed was slower compared to the capture of Figure \ref{fig:hist_rep}. The location picture of the J stairs capture set is provided in Appendix II.

 The plots display index 12 of two different datasets whose indices go up to a total of 40 which is the number of multipath signals extracted from the received signal. Here we observe the similarities in the distribution pattern of the captures. 
Although the empirical PDF are not completely matched, the general pattern is similar.  
 
 This proves that the data captured from a single location will generally follow a similar pattern in a particular environmental condition (wind, humidity, temperature, etc.), thus aiding in providing enough information to map the surface and the environment at the time of the capture. 

\begin{figure}[t] 
    \centering 
	\subfloat[][Empirical PDF near a 10 meter tall building in the presence of slow wind speed. \label{fig:hist_rep1}]{\includegraphics[width=0.49\textwidth]{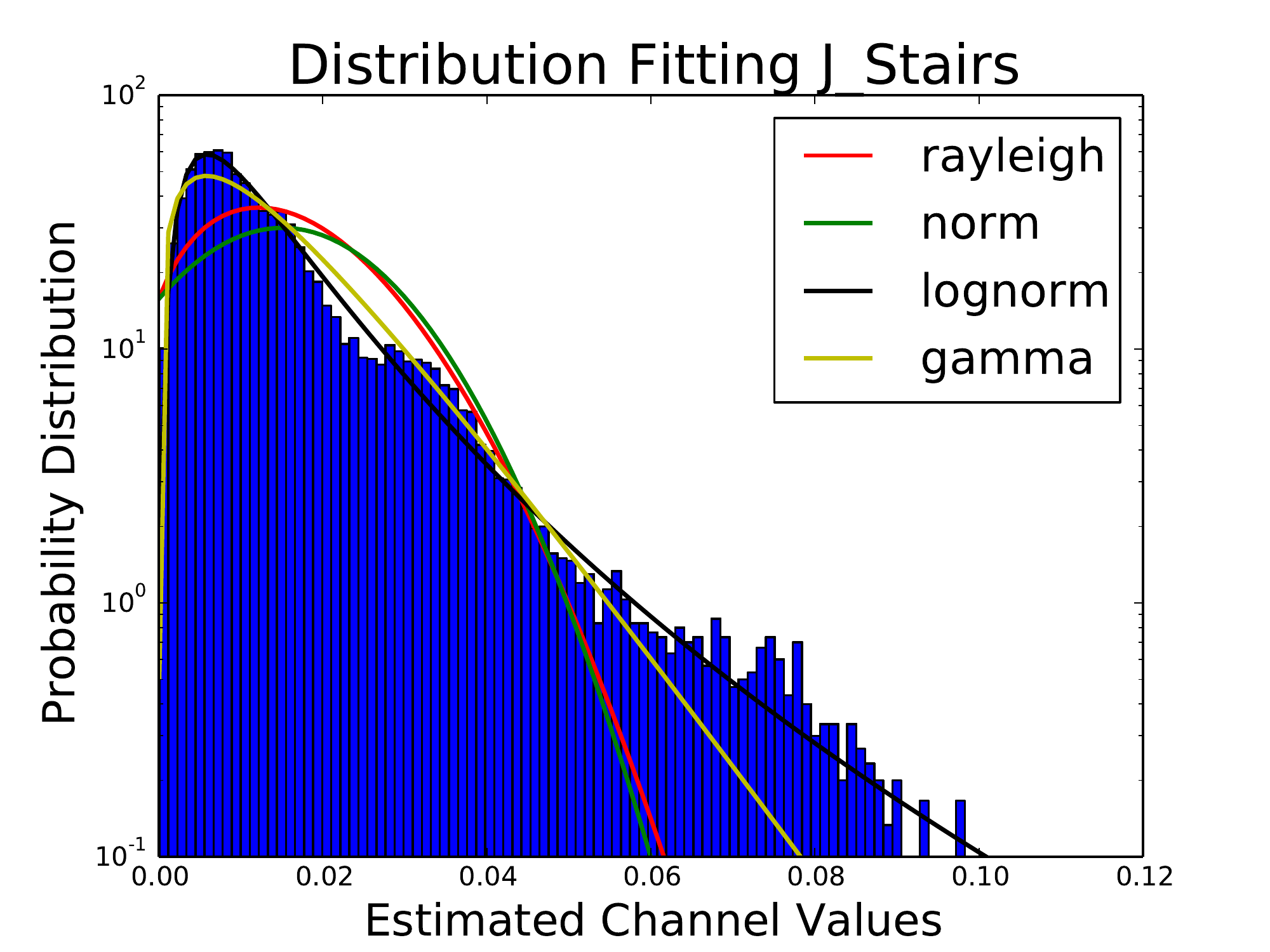}}\hfill
	\subfloat[][Empirical PDF near a 10 meter tall building when the wind speed is high. \label{fig:hist_rep}]{\includegraphics[width=0.49\textwidth]{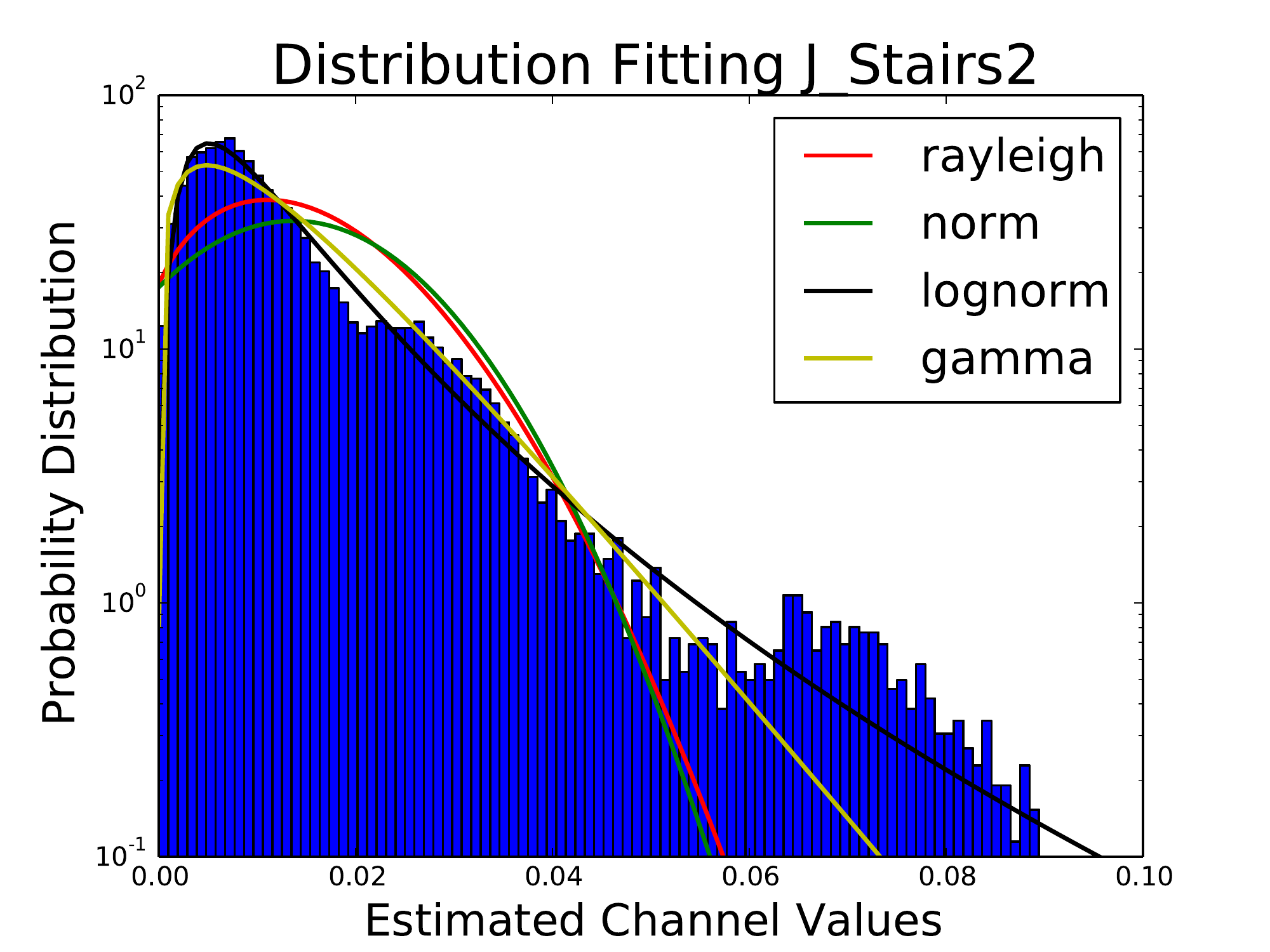}}
	\caption{Empirical PDF comparison of $12^{th}$ index taken from a 40 index dataset near a building with different wind speeds. In the legend, {\tt rayleigh}, {\tt norm}, {\tt lognorm} and {\tt gamma} represents theoretical PDF assuming Rayleigh distribution, Gaussian distribution, Lognormal distribution and Gamma distribution respectively.}
\end{figure}

\begin{figure}[h]
	\centering
	\subfloat[][Empirical PDF of captures near a bus, the receiver is located about $3$ meters from the bus. \label{fig:hist_jammie}]{\includegraphics[width=0.49\textwidth]{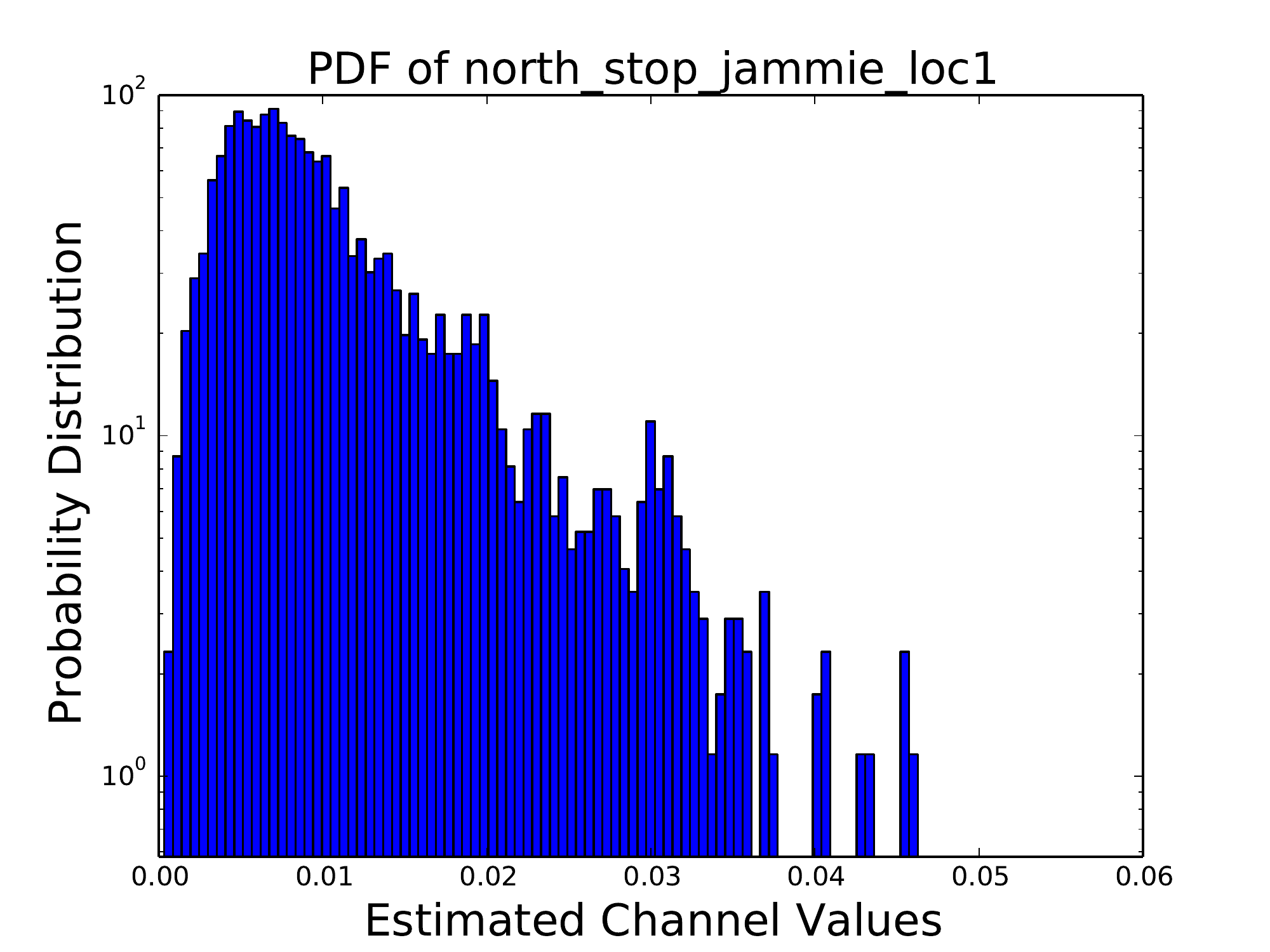}}\hfill
	\subfloat[][Empirical PDF of captures in case of single car located about $3$ meters from the receiver. \label{fig:hist_car}]{\includegraphics[width=0.49\textwidth]{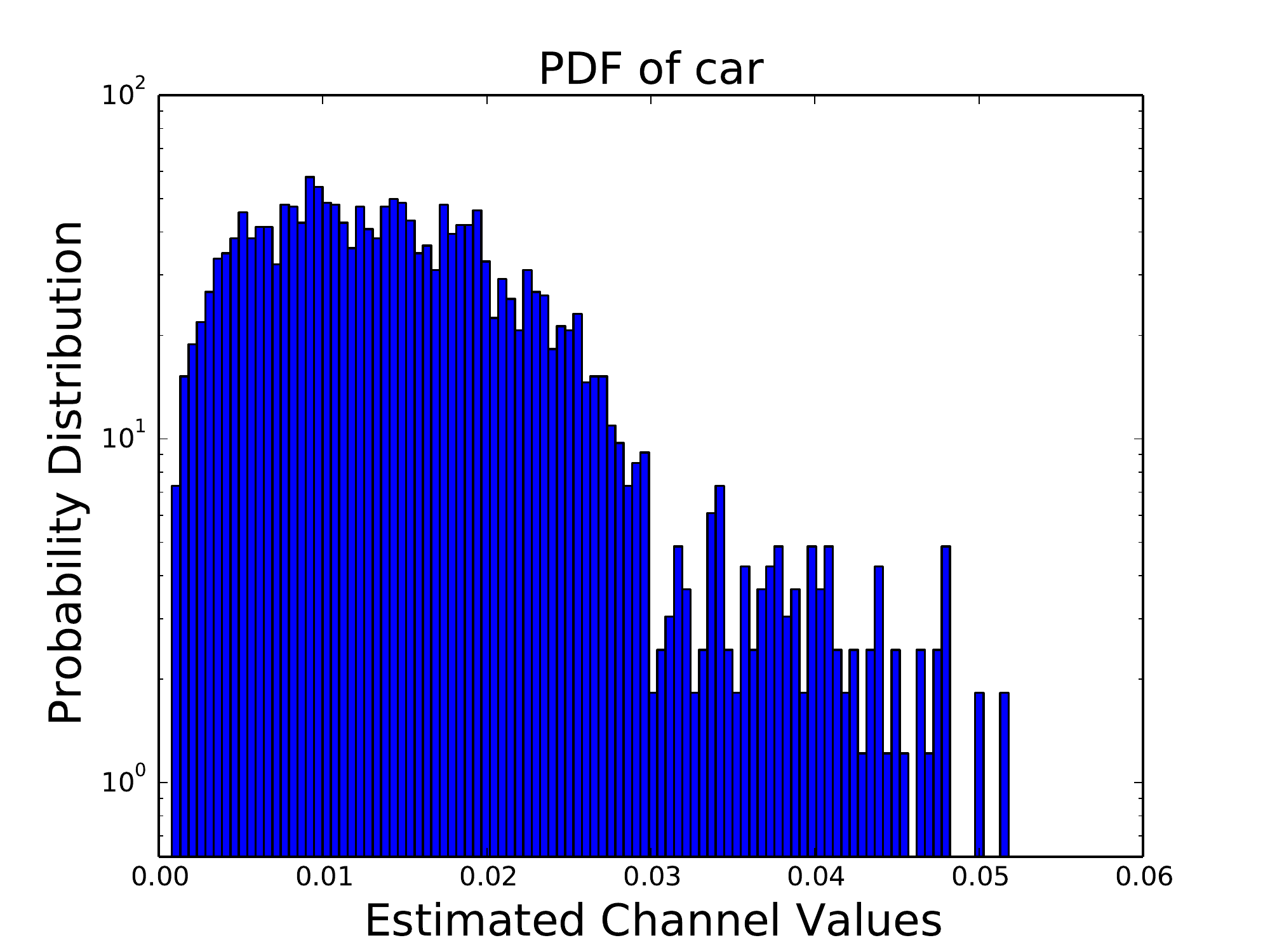}}
	\caption{Empirical PDF of the captured data in the presence of a bus and a car showing the differenecs in the distributions due to the presence of particular targets.}\label{fig:curve_fitting1}
\end{figure}

\begin{figure}[h]
	\centering
	\subfloat[][Empirical PDF of captures at a train station in the presence of a train at a distence of 6 meters from the receiver. \label{fig:hist_rts_t}]{\includegraphics[width=0.49\textwidth]{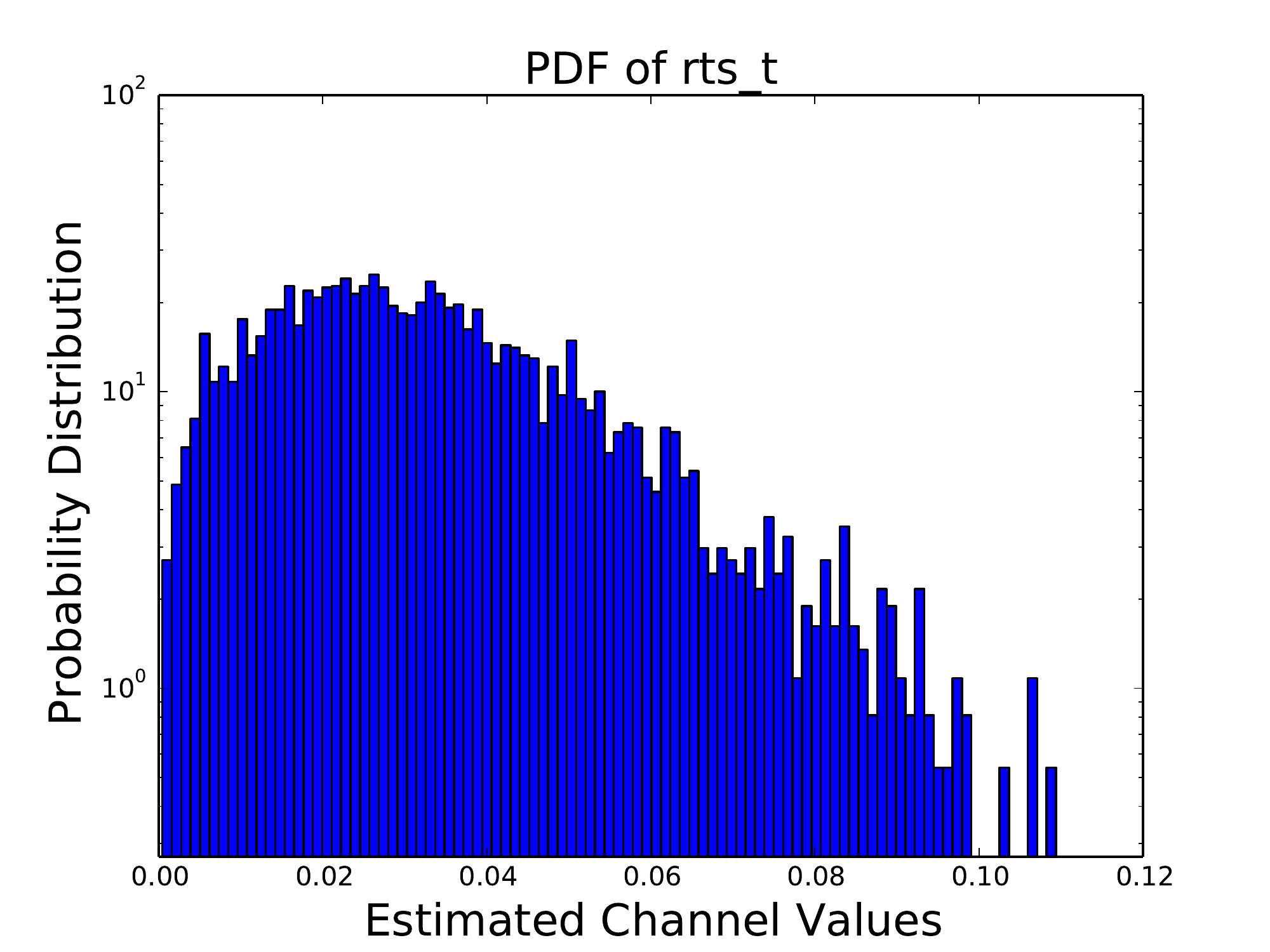}}\hfill
	\subfloat[][Empirical PDF of captures at an open air parking lot, when it is filled with cars. \label{fig:hist_parking}]{\includegraphics[width=0.49\textwidth]{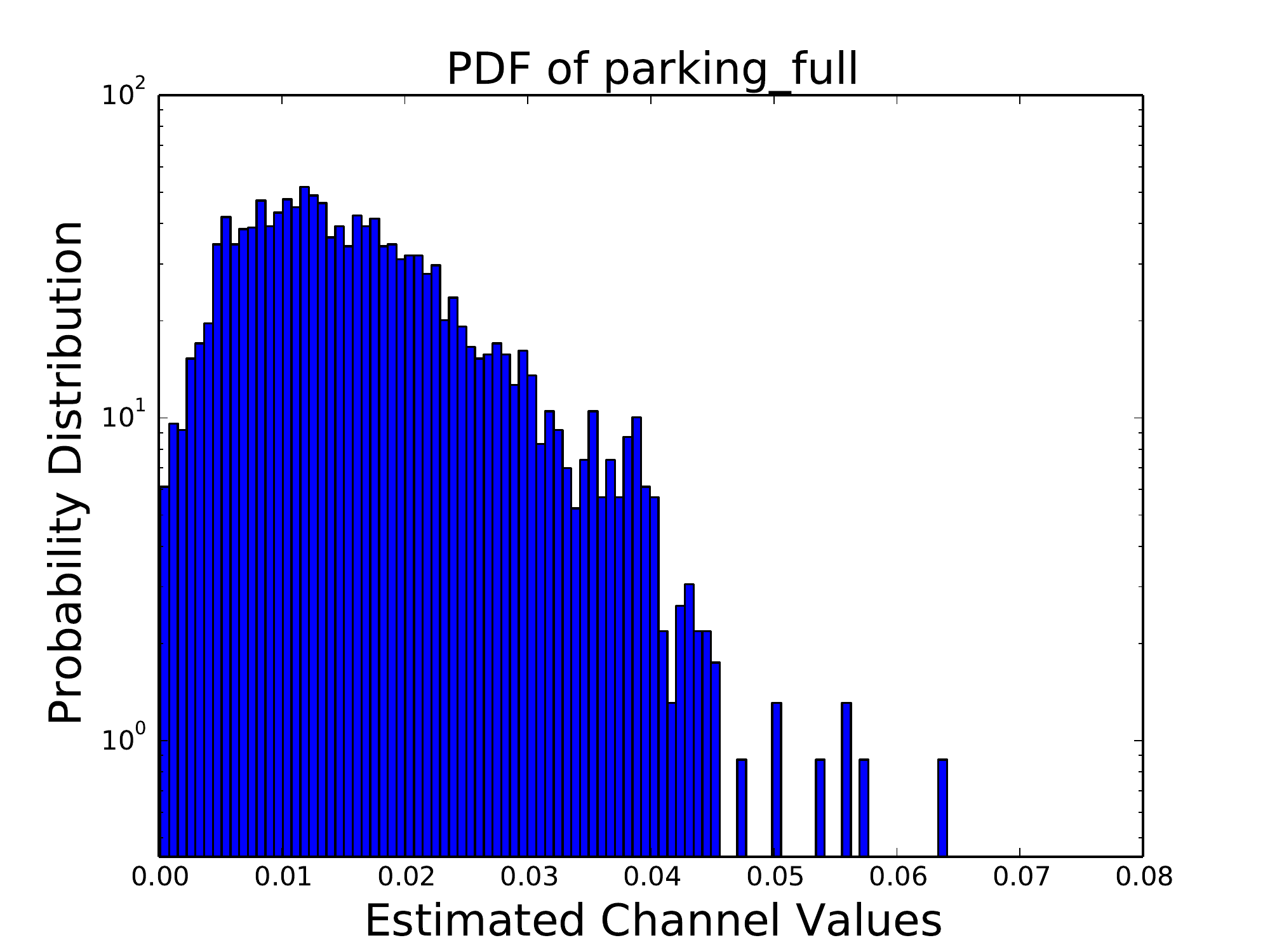}}
	\caption{Empirical PDF of captures at a full parking lot and near a train in a local train station showing the differenecs in the distributions.}\label{fig:curve_fitting2}
\end{figure}

\begin{figure}[h]
	\centering
	\subfloat[][Empirical PDF of captures taken in front of a $10$ meter tall building. The receiver is located at a distance of 8 meters from the building. \label{fig:hist_jameson}]{\includegraphics[width=0.49\textwidth]{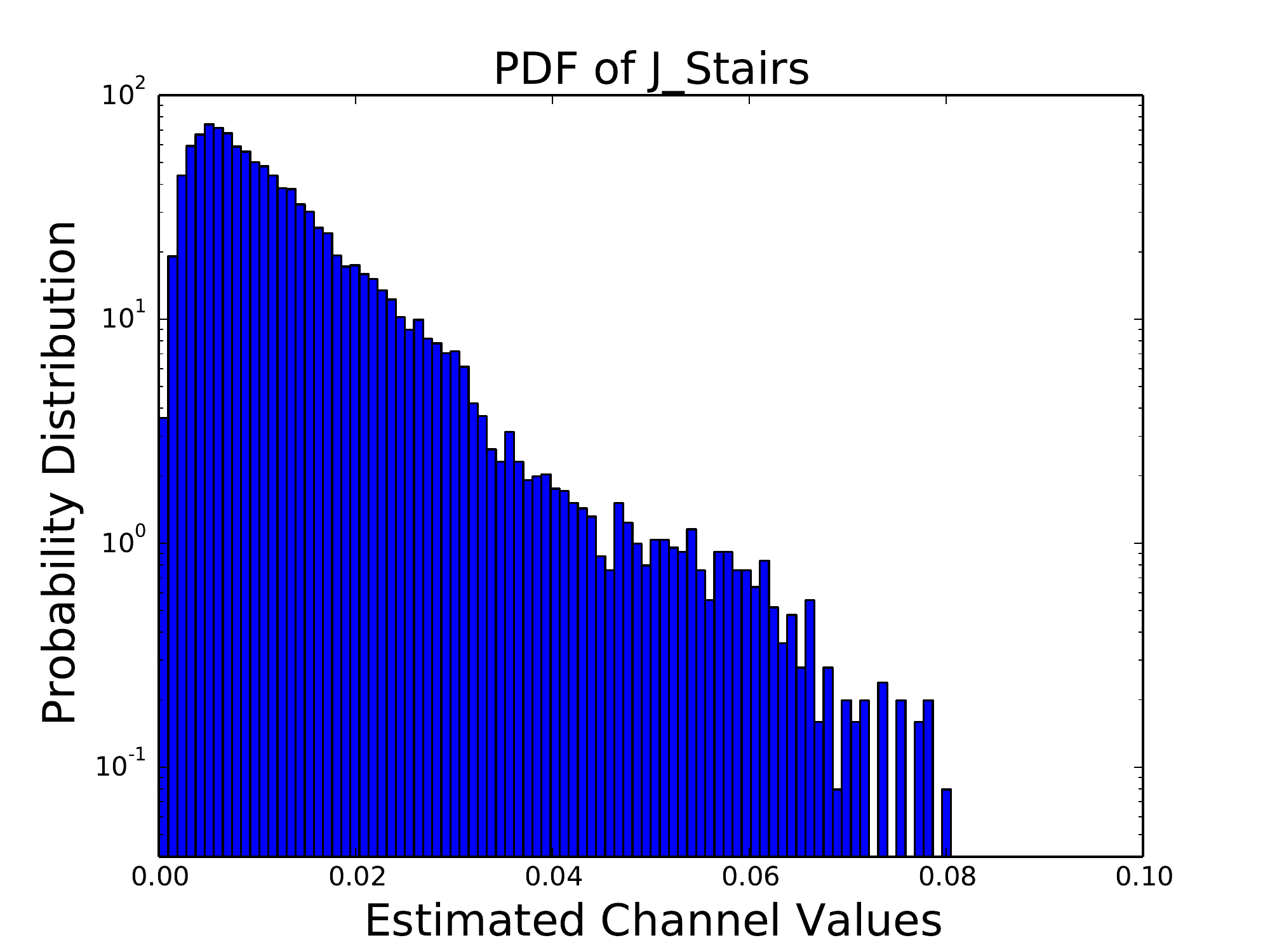}}\hfill
	\subfloat[][Empirical PDF of captures taken near a hill. \label{fig:hist_rhodesmem}]{\includegraphics[width=0.49\textwidth]{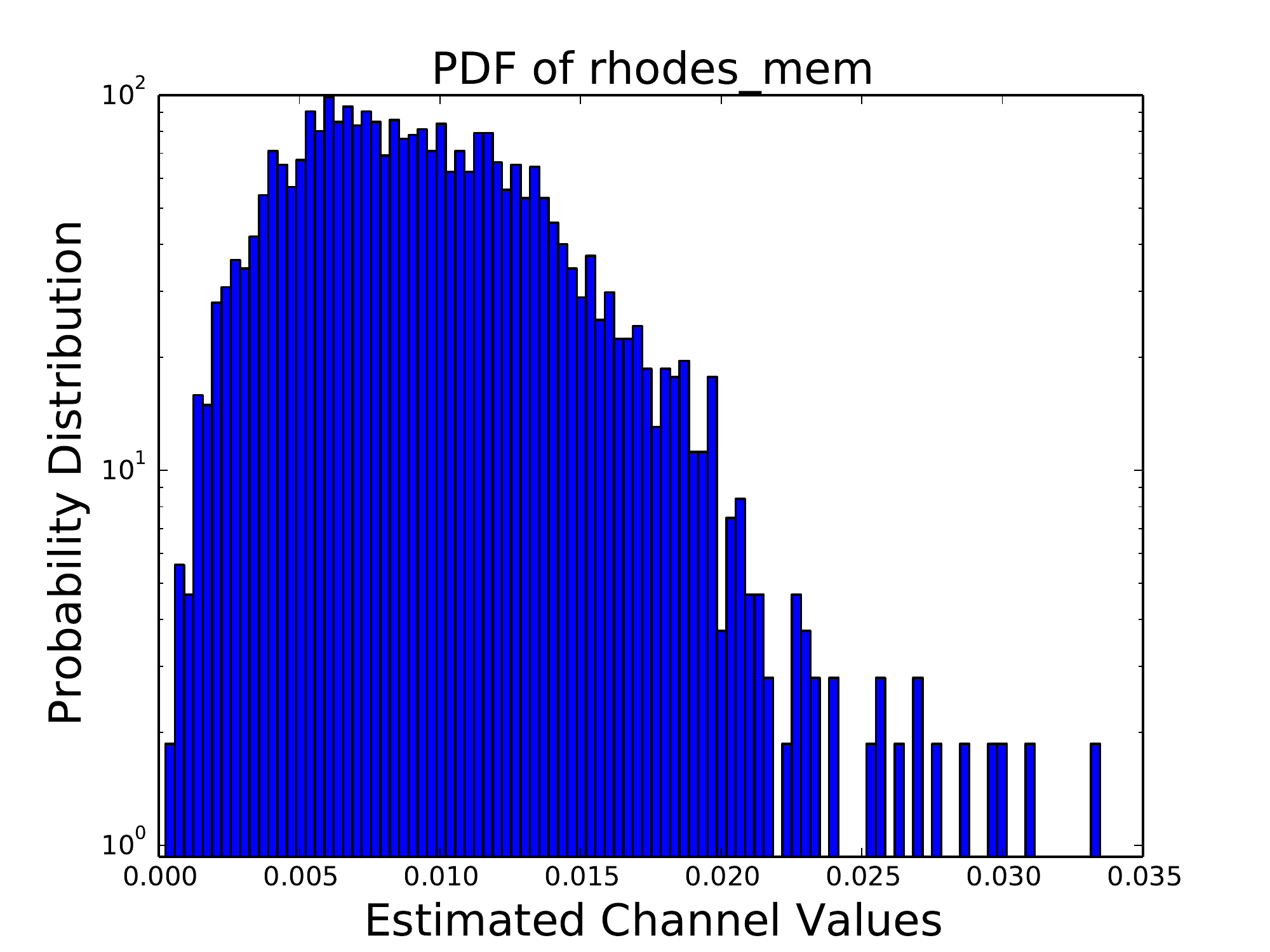}}
	\caption{Empirical PDF of captures infront of a $10$ meter tall building and near a hill. Showing the difference in the distributions between natural objects and man made objects.}\label{fig:curve_fitting3}
\end{figure}

\begin{figure}[h]
	\centering
	\subfloat[][Empirical PDF with a dihedral corner reflector. Placed in a V-H configuration, One plate was vertical and the other was kept horizontal to the ground.  \label{fig:hist_vhcr}]{\includegraphics[width=0.49\textwidth]{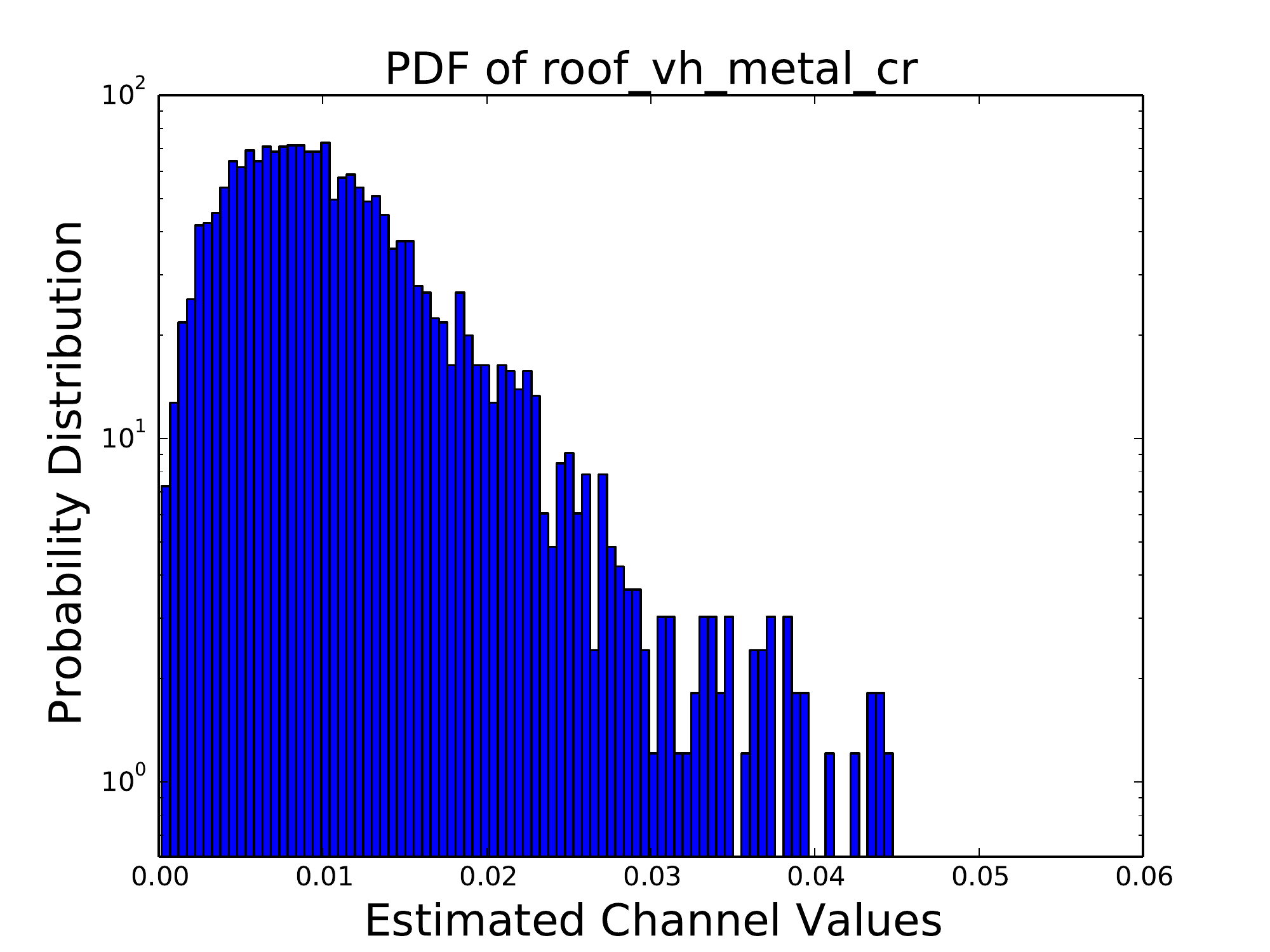}}\hfill
	\subfloat[][Empirical PDF without corner reflector. Same location and climatic conditions as Figure \ref{fig:hist_vhcr}.  \label{fig:hist_bcrl2}]{\includegraphics[width=0.49\textwidth]{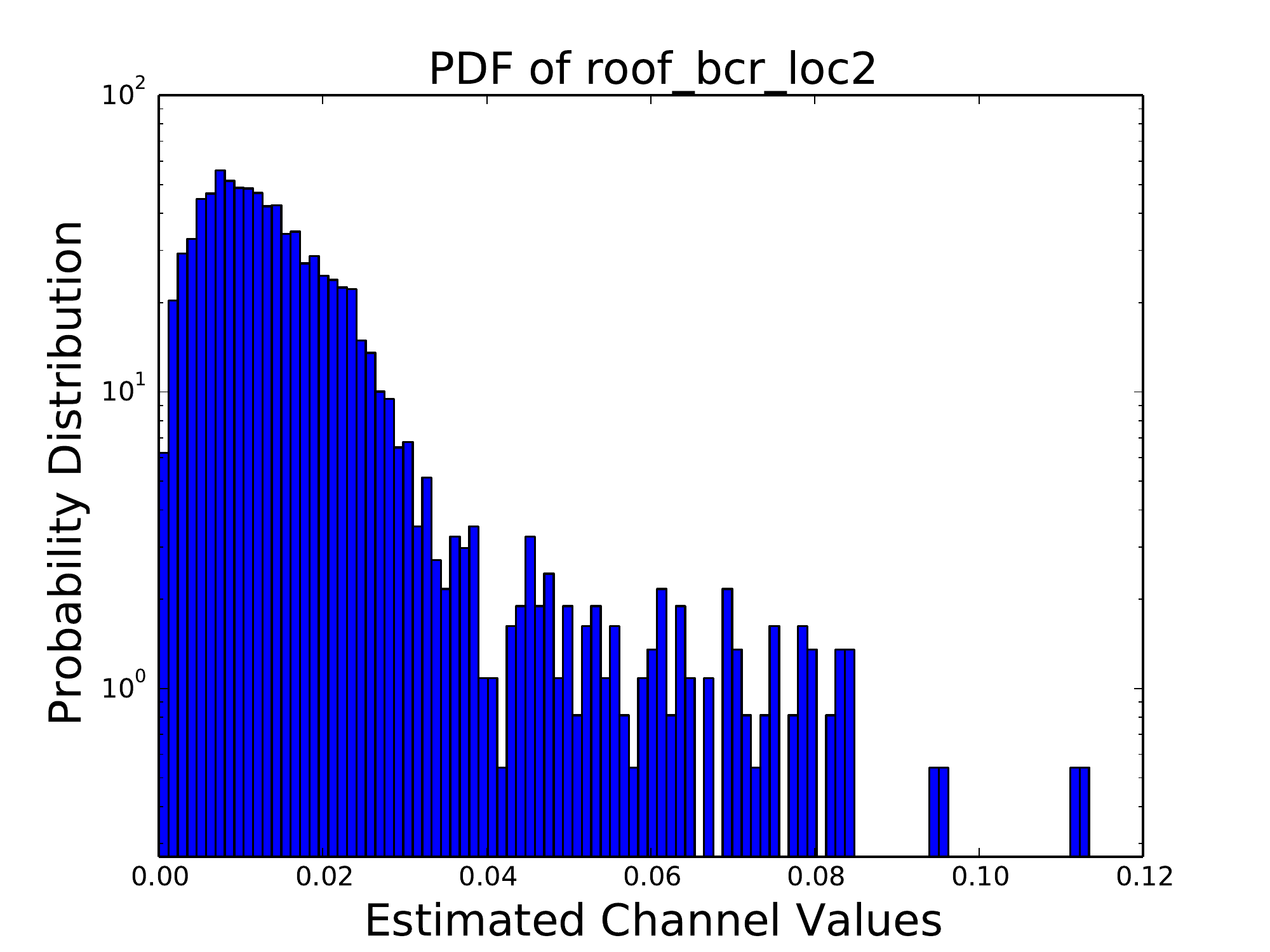}}
	\caption{Empirical PDF of captures with and without the presence of a dihedral corner reflector.}\label{fig:curve_fitting4}
\end{figure}

Next we analyse the empirical PDF and theoretical PDF of estimated channel values from different scenarios. Some of the example empirical PDFs are shown in Figures \ref{fig:curve_fitting1},\ref{fig:curve_fitting2},\ref{fig:curve_fitting3},\ref{fig:curve_fitting4}. Each curve shows the empirical PDF of a particular case. Keeping the $y$-axis constant shows the differences in the given datasets clearly. The datasets presented here include the case of car in proximity, near a big building, bus in proximity, full parking space, train in proximity and the special case with and without the corner reflector. 

\begin{table}[tbph]
\resizebox{0.95\textwidth}{!}{\begin{minipage}{1.1\textwidth}
\centering
\begin{tabular}{ | c | c | c | c | c | c | c |} \hline
\multirow{2}{*}{\textbf{Dataset}} & \multicolumn{2}{ c |}{\textbf{Shape}} & \multirow{2}{*}{\textbf{Mean}} & \multirow{2}{*}{\textbf{Variance}} & \multirow{2}{*}{\textbf{Skew}} & \multirow{2}{*}{\textbf{Kurtosis}} \\ \cline{2-3}
 & \textbf{Log-normal} & \textbf{Gamma} & & & & \\ \hline
\textbf{jameson\_stairs} & $0.762$ & $1.61$ & $1.5e^{-2}$ & $1.7e^{-4}$ & $2.15$ & $6.23$ \\ \hline
\textbf{jameson\_stairs2} & $0.767$ & $1.58$ & $1.3e^{-2}$ & $1.4e^{-4}$ & $2.38$ & $7.83$ \\ \hline
\textbf{rts\_t} & $0.28$ & $4.80$ & $4.5e^{-2}$ & $3.6e^{-4}$ & $0.58$ & $0.17$ \\ \hline
\textbf{roof\_bcr\_loc2} & $0.74$ & $1.67$ & $2.9e^{-2}$ & $6.3e^{-4}$ & $1.95$ & $4.82$ \\ \hline
\textbf{roof\_vh\_metal\_cr} & $0.46$ & $2.65$ & $1.2e^{-2}$ & $5.41e^{-5}$ & $1.16$ & $1.85$ \\ \hline
\textbf{north\_stop\_jammie\_loc1} & $0.63$ & $2.15$ & $1.2e^{-2}$ & $7.71e^{-5}$ & $1.95$ & $5.51$ \\ \hline
\textbf{car} & $0.40$ & $2.91$ & $1.5e^{-2}$ & $8.08e^{-5}$ & $0.97$ & $1.19$ \\ \hline
\textbf{rhodes\_mem} & $0.29$ & $5.55$ & $9.6e^{-3}$ & $2.43e^{-5}$ & $0.95$ & $1.76$ \\ \hline
\textbf{parking\_full} & $0.43$ & $3.07$ & $1.6e^{-2}$ & $1.02e^{-4}$ & $1.14$ & $1.92$ \\ \hline
\end{tabular}
\end{minipage}}
	\caption{Moments of captured datasets.}\label{tab:datasetmoments}
\end{table}

 In Figure \ref{fig:curve_fitting1} and \ref{fig:curve_fitting2} we observe the change in variance with change in the target conditions. In the case of a bus, the distribution has a lower variance which increases as the target changes to a car and even more in case of the train. Looking at Figure \ref{fig:hist_car} and \ref{fig:hist_parking} a similar pattern can be observed with minimum changes in the distribution. This is because both the captures represent cars in a flat space, the only difference is in Figure \ref{fig:hist_car} there is only one car in proximity and in Figure \ref{fig:hist_parking} there are many cars which creates the observable differences in the plots.
  Figure \ref{fig:curve_fitting3} gives the distribution comparison in case of two different situations, near a hill and near a tall building. The variance in case of the hill is a lot less compared to the variance in case of the building thus differentiating between them. Figure \ref{fig:curve_fitting4} is a special case where the captures are taken with a corner reflector placed 2 meter from the receiver antenna and the differences in the plots are clearly visible. Table~\ref{tab:datasetmoments} provides the various moments extracted directly from the captured data. The shape parameters are extracted by fitting the empirical data to the expected distribution, thus there is a difference between the log-normal and gamma distribution shape parameters. These correlates with the information visible in the plots and gives us a better understanding of the changes in parameters due to the change in environmental conditions.
  
  From the plots and table provided here it is observable that there are differences in the distributions of the channel information due to the changes in the physical environment. This enables us with the opportunity to utilise these differences and model a system that can differentiate between them. Although the differences are quiet clear from the empirical PDF's, it is always beneficial to further analyse the received data and gain better understanding.

\subsection{Chi-square Test} \label{subsec:chisquare}
 The second set of analysis performed on the captured dataset is the Chi-square test \cite{vardasbi2011multiple,kendall1961advanced}. This is a test which takes two inputs, the observed data and the expected data and gives out a score based of the two following the equation (\ref{eq:chisq}). The Degree of Freedom (DF) in this case is 100 since the length of the array is 101 and DF is defined as one less than the length of the array.

\begin{equation}\label{eq:chisq}
\chi^{2} = \sum_{i=1}^{M}\dfrac{(O_i - E_i)^2}{E_i}
\end{equation}

Here $O_i$ is the observed value and $E_i$ is the expected value. In order to perform this analysis PDF of the received data (observed data) and the distribution fitted data (expected data) are passed through equation (\ref{eq:chisq}) and the output of the test is recorded in Table~\ref{tab:chisquare}. The p-value is defined as the probability of obtaining a specific value equal to or higher than what actually is observed, maintaining the assumption that the model is correct. The points where the p-value is null can be rejected because this means the probability of getting a value similar or greater than that is null. All other cases in this test can be accepted and Table~\ref{tab:chisquare} shows that Log-normal distribution has the most consistent result. In the next subsection we reduce the dimensionality of the data and gain better visualization by rotating the axes.

\begin{table}[tbph]
\resizebox{0.8\textwidth}{!}{\begin{minipage}{1.2\textwidth}
\centering
\begin{tabular}{ | c | c | c | c | c | c | c | c | c |} \hline
\multicolumn{9}{|c|}{\textbf{Chi-Square Value}} \\ \hline
\multirow{2}{*}{\textbf{Capture Set Names}} & \multicolumn{2}{c|}{\textbf{Rayleigh}}  & \multicolumn{2}{c|}{\textbf{Gaussian}} & \multicolumn{2}{c|}{\textbf{Log-normal}} & \multicolumn{2}{c|}{\textbf{Gamma}}\\ \cline{2-9}
 & Chi-Square & p-value & Chi-Square & p-value & Chi-Square & p-value & Chi-Square & p-value \\ \hline
\textbf{jameson\_stairs} & $752438.54$  & $0.0$  & $87119627.61$ & $0.0$ & $447.26$ & $3.34e^{-65}$ & $452.97$ & $2.61e^{-66}$\\ \hline
\textbf{jameson\_stairs2} & $199326.03$ & $0.0$ & $11569636.34$ & $0.0$ & $514.14$ & $2.79e^{-78}$ & $513.09$ & $4.48e^{-78}$\\ \hline
\textbf{rts\_t} & $352.03$ & $5.47e^{-47}$ & $9127.03$ & $0.0$ & $280.29$ & $8.73e^{-34}$ & $281.76$ & $4.75e^{-34}$\\ \hline
\textbf{roof\_bcr\_loc2} & $20128.95$  & $0.0$ & $529926.62$ & $0.0$ & $265.25$ & $4.34e^{-31}$ & $275.86$ & $5.47e^{-33}$\\ \hline
\textbf{roof\_vh\_metal\_cr} & $977.26$  & $3.59e^{-172}$ & $1100.82$ & $9.18e^{-198}$ & $970.27$ & $9.97e^{-171}$ & $972.30$ & $ 3.78e^{-171}$\\ \hline
\textbf{north\_stop\_jammie\_loc1} & $2710.12$  & $0.0$ & $104025.97$ & $0.0$ & $734.89$ & $1.67e^{-122}$ & $735.07$ & $1.53e^{-122}$\\ \hline
\textbf{car} &  $975.44$  & $8.53e^{-172}$ & $931.99$ & $7.79e^{-163}$ & $969.07$ & $1.76e^{170}$ & $ 973.73$ & $1.91e^{-170}$\\ \hline
\textbf{rhodes\_mem} &  $1502.04$  & $1.18e^{-281}$ & $1470.23$ & $5.70e^{-275}$ & $1499.21$ & $4.64e^{-281}$ & $1501.06$ & $1.89e^{-281}$\\ \hline
\textbf{parking\_full} &  $700.52$  & $1.54e^{-115}$ & $1877.42$ & $0.0$ & $679.31$ & $2.97e^{-111}$ & $680.71$ & $1.55e^{-111}$\\ \hline
\end{tabular}
\end{minipage}}
	\caption{Chi-Square test values for the received data matched with the fitted data for each distribution.}\label{tab:chisquare}
\end{table}

\subsection{Principle Component Analysis (PCA)} \label{subsec:PCA}
In this subsection we investigate the clustering of the estimated channel values in different scenarios. In order to better visualize different sets of the captured data we used PCA. 
 With this analysis we can reduce the dimensionality of the data and view the datasets from an angle that provides maximum information. Here we have observed different cluster formation of the datasets due to the change in locations and environmental conditions. There are many different ways to reduce the dimensionality of the data and calculate the principle components, for this implementation we have used Singular Value Decomposition (SVD) \cite{abdi2007eigen,abdi2007singular,takane2003relationships}. The major goals of PCA are to reduce the dimensionality of the data, extract the important information and analyse the structure of the data. The calculations for PCA are shown in \cite{abdi2010principal}.
  
 In order to derive the principle components from the data we first need to generate the SVD equivalent of the dataset \textbf{A}. The input matrix \textbf{A} has $J$ sets of data explained by $K$ variables, represented by $J\times K$. \textbf{A} has a rank $L$ with $L \leq \min \{J,K\}$, then the SVD of \textbf{A} will be given by equation (\ref{eq:svd}).
 
\begin{align}
\textbf{A} &= \textbf{U} \textbf{$\Delta$} \textbf{V}^T \label{eq:svd}\\
\textbf{F} &= \textbf{U} \textbf{$\Delta$} \label{eq:PCA1}
\end{align}

Here \textbf{U} is $1\times L$ matrix of left singular vectors, \textbf{V} is $K\times L$ matrix of right singular vectors and \textbf{$\Delta$} is a diagonal matrix of singular values 
The components for PCA are obtained from the data \textbf{A} using equation (\ref{eq:svd}). 
The principal component matrix \textbf{F} of dimension $J \times L$ is given by equation (\ref{eq:PCA1}). 
To get the coefficients of the linear combinations which are used to compute the factor scores the matrix \textbf{V} is used. The matrix can also be interpreted as the projection matrix because $\textbf{A}\ \text{times} \ \textbf{V} $ gives the projection values of the observations on the principle components as shown in equation (\ref{eq:PCA2}). 
\begin{equation}\label{eq:PCA2}
\textbf{F} = \textbf{U}\textbf{$\Delta$} = \textbf{U}\textbf{$\Delta$} \textbf{V}^T \textbf{V} = \textbf{A}\textbf{V} 
\end{equation}
 
 Geometrically the components can also be represented by rotating the original axes and the matrix \textbf{A} can be interpreted as a product of the factor scores given by equation (\ref{eq:PCA3}).\cite{abdi2010principal} Here \textbf{I} is the identity matrix.
\begin{equation}\label{eq:PCA3}
\textbf{A} = \textbf{F}\textbf{V}^T \qquad \textbf{F}^T \textbf{F} = \textbf{$\Delta$}^2 \quad \& \quad \textbf{V}^T \textbf{V} = \textbf{I}
\end{equation}

\begin{figure}
	\centering
	\includegraphics[width=0.8\textwidth]{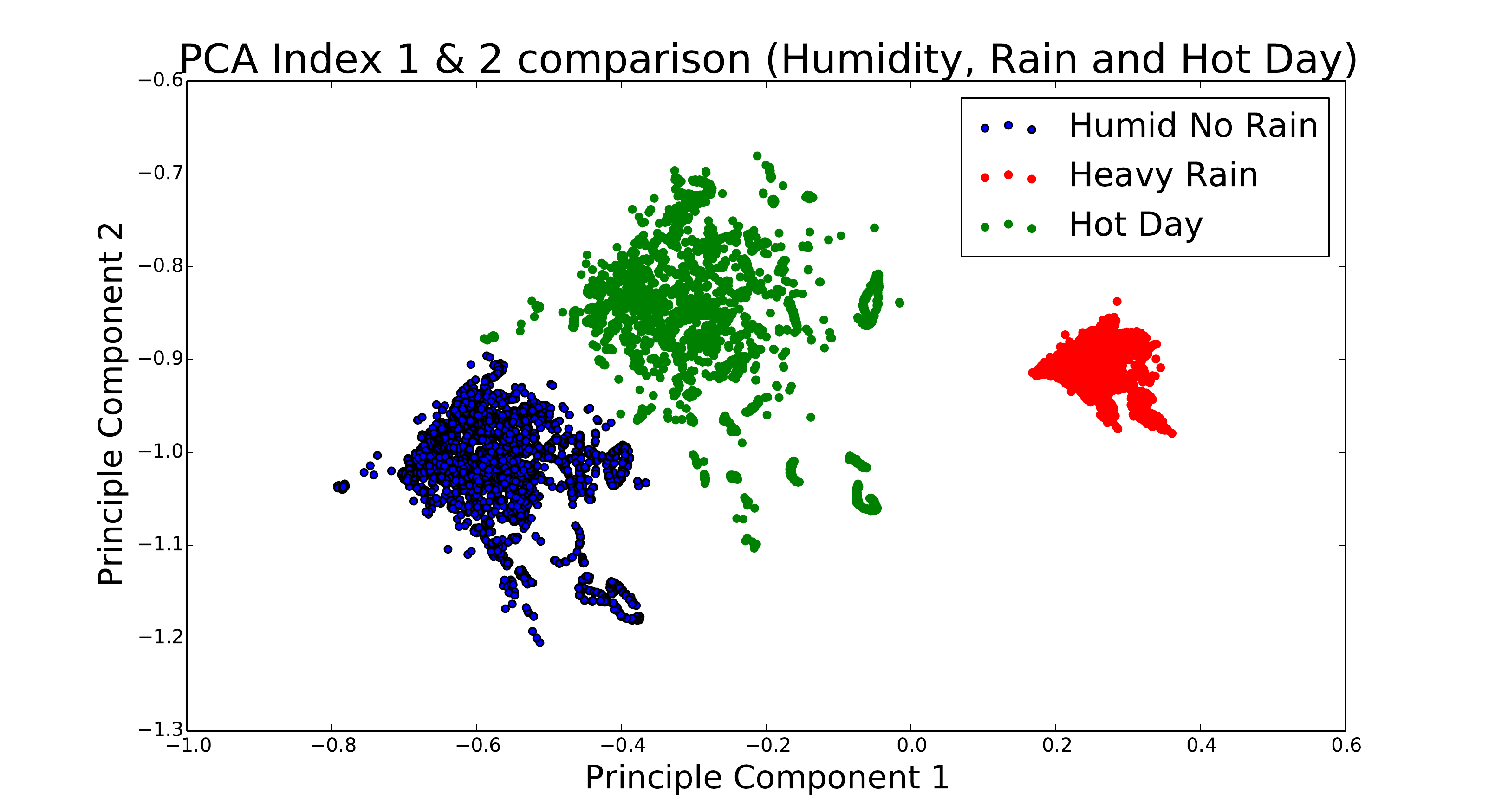}
	\caption{PCA of different climatic conditions at a single location. The conditions include high humidity but no rain, heavy rain and Hot day with very low humidity.} \label{fig:PCALoc5}
\end{figure}
 
\begin{figure}
    \centering 
    \subfloat[][The clusters shown here include the captures taken near a beach, highland, building and train. \label{fig:PCALoc1}]{\includegraphics[width=0.49\textwidth]{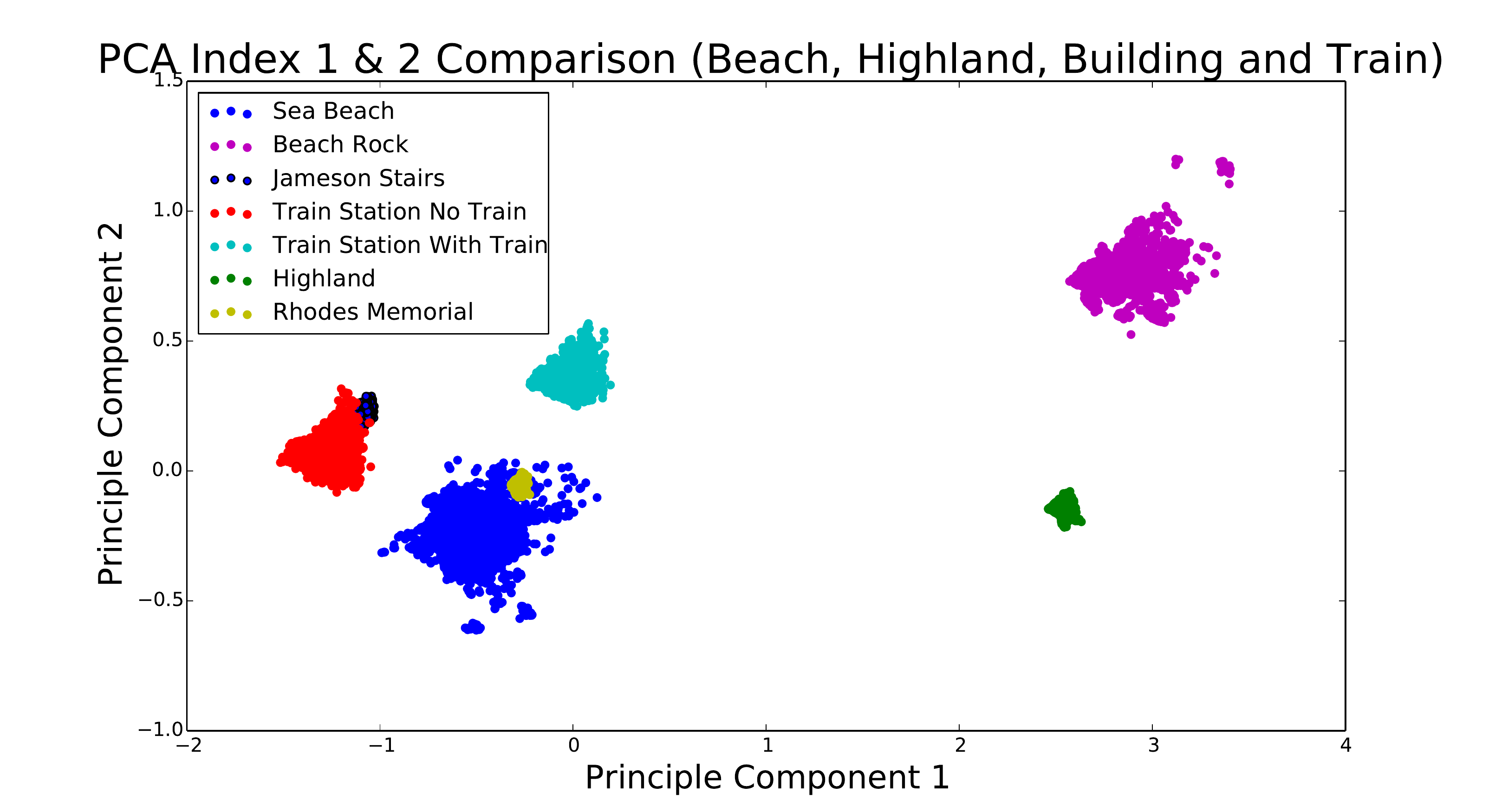}}\hfill
	\subfloat[][The clusters here are taken with different configuration of Corner Reflector placed at a distance of 2 meters from the receiver.\label{fig:PCALoc2}]{\includegraphics[width=0.49\textwidth]{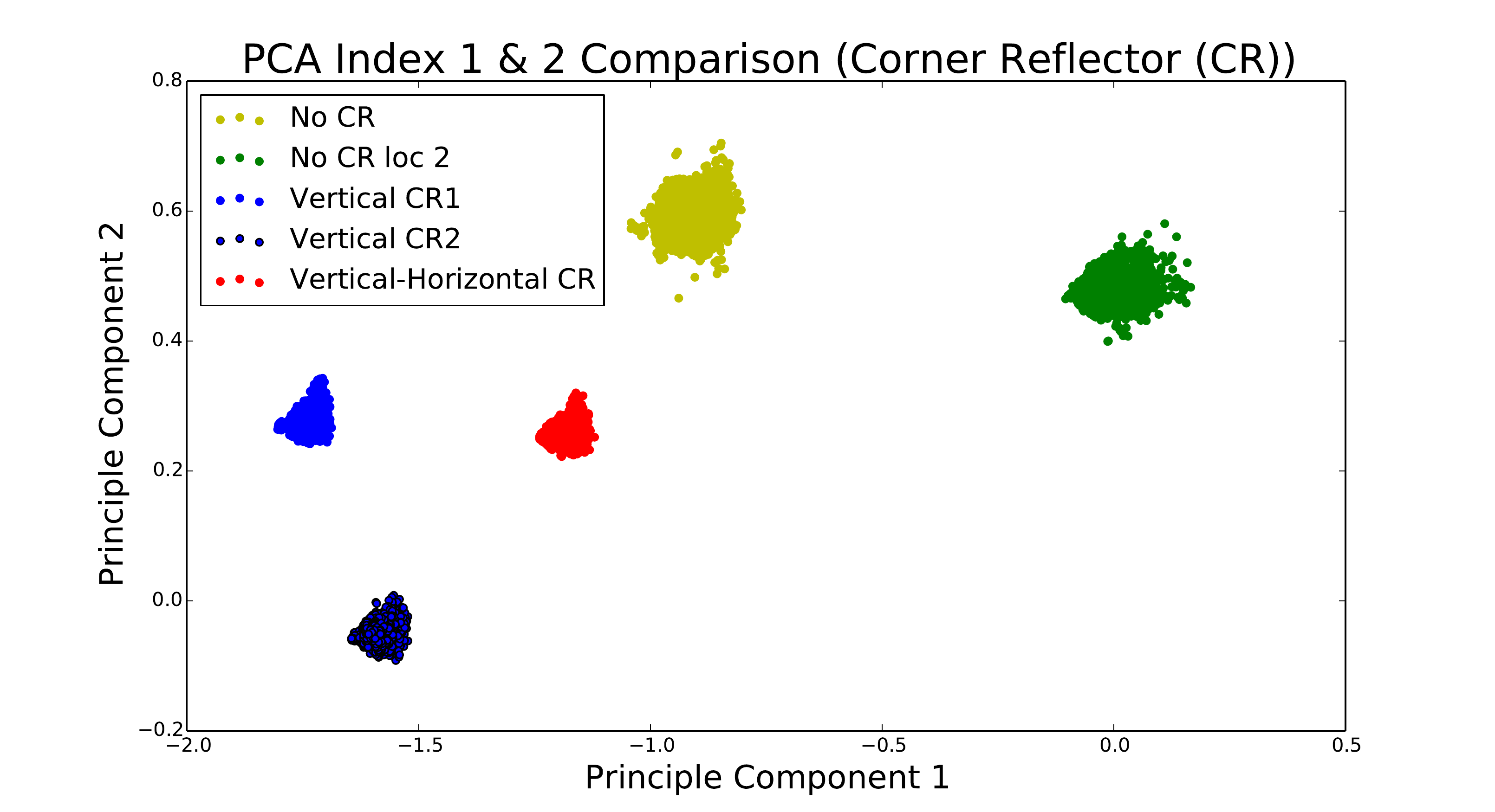}}\hfill
	\subfloat[][Different clusters formed by captures taken near a bus in different configurations and locations.\label{fig:PCALoc3}]{\includegraphics[width=0.49\textwidth]{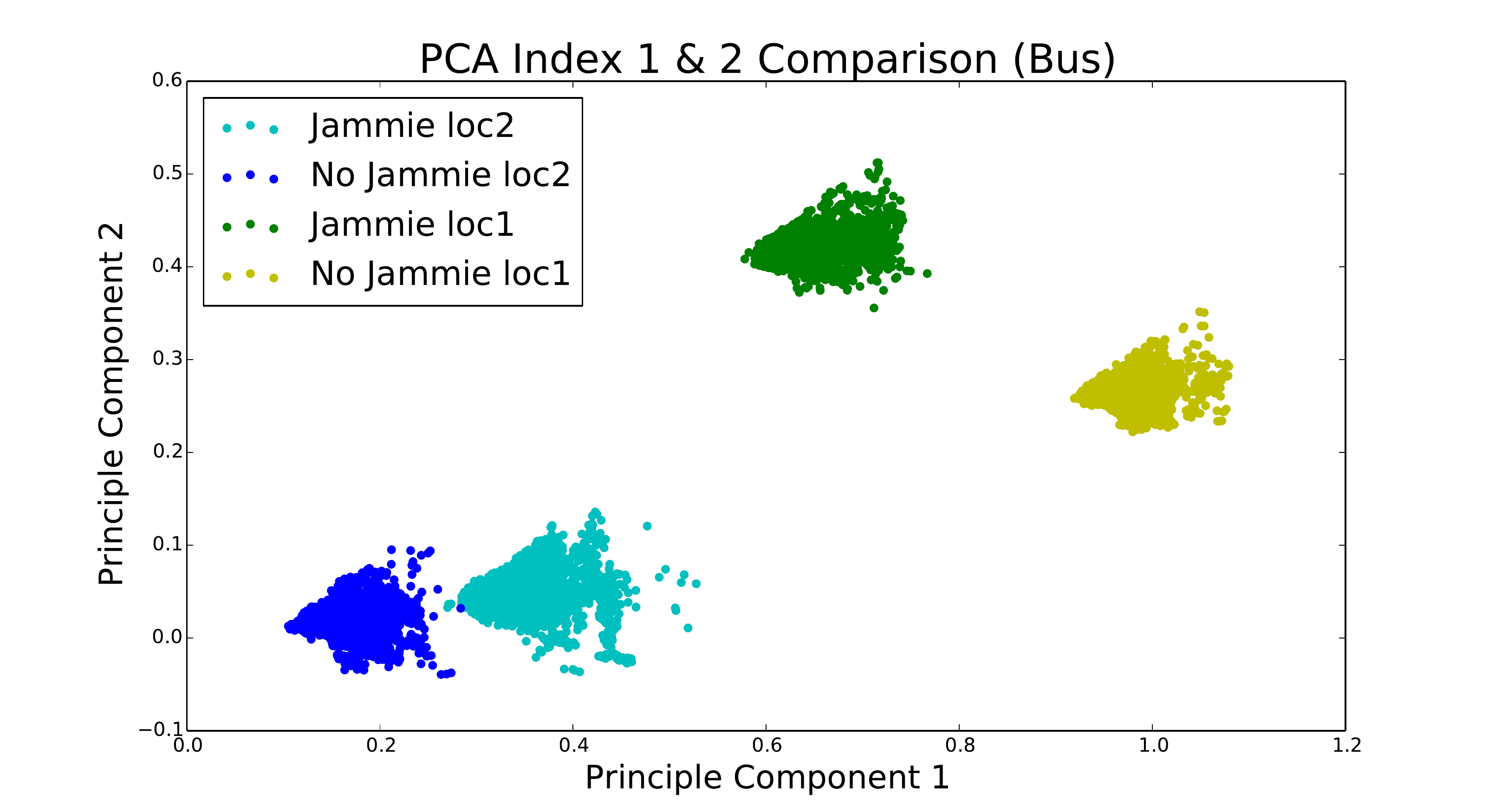}}\hfill
	\subfloat[][Clusters of captures taken at an open air parking lot when it was full/empty and with a car in proximity. \label{fig:PCALoc4}]{\includegraphics[width=0.49\textwidth]{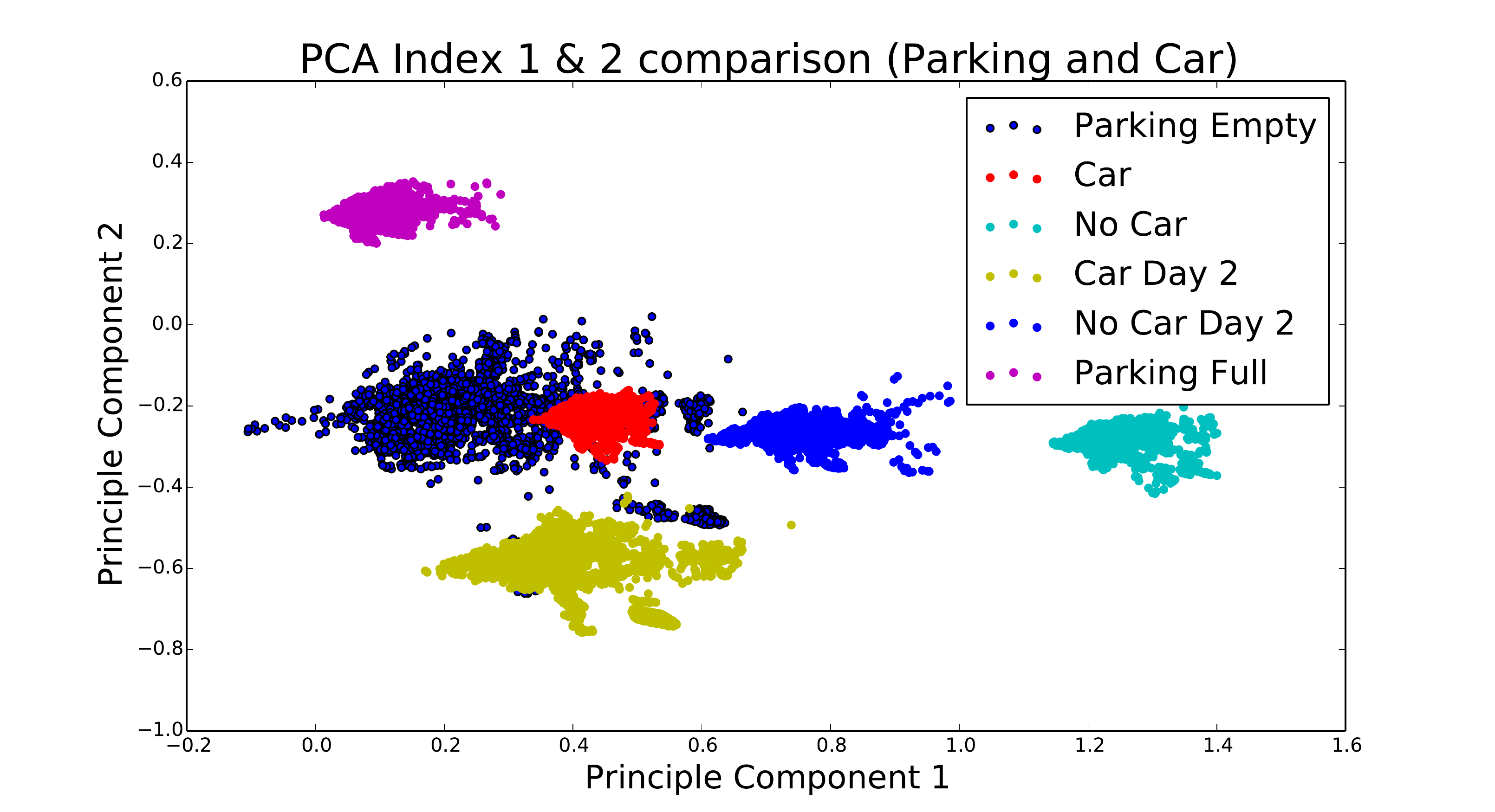}}
	\caption{PCA of different datasets showing the different clusters.}
\end{figure}

 All the plots after PCA are done by stacking the normalized datasets together and comparing the $1^{st} \text{ and the } 2^{nd}$ principle components. 
 The dataset dimensions have been reduced from 40 components to 2 major components and the relationship between the components are plotted to show the best view. The clusters are clearly visible for each data. It can be noted here that taking more components will get us more information and hence will make scene classification even easier.

 Figure \ref{fig:PCALoc1} shows the the clusters of the data captured at sea beach, near the shore, beach rock, at the shore but at a height of around 10 meters from the sea level. J stairs, stairs at the entrance of a building about 10 meters tall. Train station, gives the captures taken at a local train station in two different situations first, when there is a train about 6 meters from the receiver and secondly in the absence of the train. Highland, refers to a location where the captures are taken on the road next to a small hilly terrain. Rhodes memorial, is situated mid way up on to a mountain in Cape Town.

 Figure \ref{fig:PCALoc2} shows a particular set of captures taken with a corner reflector. The different clusters in this figure shows the different set of captures taken at various different configurations. CR is an acronym for corner reflector the different allocations are in this plot. Figure \ref{fig:PCALoc3} shows the clusters for specific set of captures taken at a bus stop on the campus of University of Cape Town (UCT) at two different locations. The university shuttle service is known as "jammie", thus the name given. The different scenarios for this capture are, in the presence of a jammie about 4 meters away from the receiver antenna in location 1, 2 and in the same locations when there is no jammie.

 Figure \ref{fig:PCALoc4} shows the captures taken at a flat parking space when there are no cars and in the presence of cars. This set also shows the plots for a single car dataset, where the captures are taken with and without a car in the same location on two different days (one at the morning 8 am and the other in the evening 6 pm) with different climatic conditions. Figure \ref{fig:PCALoc5} shows the difference in the clusters at the same location when there is heavy rain, medium rain and on a hot day.

\section{Conclusion}\label{sec:conc}
In this work we present a novel way of sensing the environment using communication signals, which we call CommSense. 
The hypothesis under test was that we will be able to sense the environment by analysing the channel equalization values of any communication system. 
To prove our hypothesis we chose GSM communication. 
We have implemented a very basic correlation based techniques to estimate the channel values from the received signal using the known training sequence. 
In the limited drive to prove the hypothesis we collected data from different scenarios. 
On analysis it was shown that the probability distribution of the collected data differs as the environment changes. 
Hence a classic Hypothesis-test type algorithm can distinguish the scenes. 
Secondly, the data for different scenarios are highly clustered. This shows that a naive nearest neighbour algorithm can distinguish the environmental conditions easily. With these we prove our hypothesis. In the future work we intend to gather more data and characterize the environment for robust classification. We also intend to work on increasing the number of receiver nodes as another way towards getting robust environment sensing using CommSense.

\section*{Appendix I}\label{Appendix I}
In this appendix we shall briefly describe four of the most used PDF models. 

\subsection{Normal Distribution (Gaussian Distribution)}

 The most general form of distribution used in statistical analysis is the gaussian distribution. The probability density function follows the equation (\ref{eq:gauss}).
 
\begin{equation}\label{eq:gauss}
f(x) = \frac{1}{\sigma \sqrt{2\pi }} e^{\frac{-(x-\mu)^2 }{2\sigma^2}}
\end{equation}
where, \\
$\mu$ = location parameter (mean)\\
$\sigma$ = scale parameter (standard deviation)\\

Standard Normal Distribution where $\mu = 0 \ \& \ \sigma = 1$ is given as equation (\ref{eq:std_normal})
\begin{equation}\label{eq:std_normal}
f(x) = \frac{1}{\sqrt{2\pi}}e^{\frac{-x^2}{2}}
\end{equation}

The maximum likelihood estimation for the gaussian distribution is derived by minimizing the log-likelihood function of the equation (\ref{eq:gauss}) given by equation (\ref{eq:mle_gauss}),(\ref{eq:mle_gauss1})
\begin{align}
\hat{\mu} &= \frac{1}{n} \sum_{i=1}^{N} x_i \label{eq:mle_gauss}\\
\hat{\sigma}^2 &= \frac{1}{n} \sum_{i=1}^{N} (x_i - \hat{\mu})^2\label{eq:mle_gauss1}
\end{align}

\subsection{Rayleigh Distribution}

 One of the most common distributions used in wireless communication systems to model the channel is Rayleigh distribution. The probability density function is given by the equation (\ref{eq:rayleigh}).
 
\begin{equation}\label{eq:rayleigh}
f(x) =  \frac{1}{\sigma^2} x e^{\frac{-x^2}{2\sigma^2}}
\end{equation}
where, \\
$\sigma$ = scale parameter (mode)\\

The maximum likelihood estimation for the parameters of rayleigh distribution is given by equation (\ref{eq:mle_rayleigh})
\begin{equation}
\hat{\sigma}^2 =\frac{1}{2n}\sum_{i=1}^{N}x_i^2 \qquad x>0 \label{eq:mle_rayleigh}
\end{equation}

\subsection{Gamma Distribution}
 Gamma distribution is also a very versatile distribution which can be manipulated in many ways to give other distributions such as k-distribution which is in the scope of the future work. The general formula for the probability density function of gamma distribution is given by equation (\ref{eq:gamma})

\begin{equation}\label{eq:gamma}
f(x) = \frac{(\frac{x-\mu}{\beta})^{\gamma - 1}\exp{(-\frac{x-\mu}{\beta}})} {\beta\Gamma(\gamma)}  \qquad  x \ge \mu; \gamma,\beta > 0
\end{equation}
Where, \\
$\gamma$ = shape parameter\\
$\mu$ = location parameter\\
$\beta$ = scale parameter\\
$\Gamma$ is the gamma function given by equation (\ref{eq:gamma_func})
\begin{equation}\label{eq:gamma_func}
\Gamma(a) = \int_{0}^{\infty} {t^{a-1}e^{-t}dt}
\end{equation}

 The case where $\mu = 0 \ \& \ \beta = 1 $ is called the standard gamma distribution and is given by equation (\ref{eq:std_gamma})
\begin{equation}\label{eq:std_gamma}
f(x) = \frac{x^{\gamma - 1}e^{-x}} {\Gamma(\gamma)}  \qquad x \ge 0; \gamma > 0
\end{equation}

 The maximum likelihood estimates for the two parameter gamma distribution are calculated by solving the following equations (\ref{eq:mle_gamma}) and (\ref{eq:mle_gamma1}) simultaneously 
\begin{align}
\hat{\beta} - &\frac{\bar{x}}{\hat{\gamma}} = 0 \label{eq:mle_gamma}\\
\log{\hat{\gamma}} - \psi(\hat{\gamma}) - &\log \left( \frac{\bar{x}}{ \left( \prod_{i=1}^{n}{x_i} \right) ^{1/n}  } \right) = 0 \label{eq:mle_gamma1}
\end{align}
 with $\psi$ denoting the digamma function  given by equation (\ref{eq:digamma}), which is the mathematical derivative of the gamma function. These functions are solved by using python's scikit learn package.
\begin{equation}\label{eq:digamma}
\psi(z) \equiv \frac{d}{dz} \ln \Gamma(z) = \frac{\Gamma^{'}(z)}{\Gamma(z)}
\end{equation}

\begin{table}[tbph]
\resizebox{\textwidth}{!}{\begin{minipage}{\textwidth}
\centering
\begin{tabular}{ | c | c | c | c | c | c |} \hline
\textbf{Distribution} & \textbf{Shape} & \textbf{Mean} & \textbf{Variance} & \textbf{Skew} & \textbf{Kurtosis} \\ \hline
\textbf{Gaussian} & N/A & $0.0$ & $1.0$ & $0.0$ & $0.0$\\ \hline
\textbf{Rayleigh} & N/A & $1.25$ & $0.43$ & $0.63$ & $0.25$  \\ \hline
\multirow{4}{*}{\textbf{Log-normal}} & $0.2$ & $1.02$ & $0.04$ & $0.61$ & $0.68$ \\ \cline{2-6}
& $0.5$ & $1.13$ & $0.36$ & $1.75$ & $5.89$ \\ \cline{2-6}
& $1.0$ & $1.64$ & $4.67$ & $6.18$ & $110.93$ \\ \cline{2-6}
& $2.0$ & $7.39$ & $ 2926.35$ & $414.36$ & $9220556.98$ \\ \hline
\multirow{4}{*}{\textbf{Gamma}} & $0.5$ & $0.5 $ & $0.5 $ & $2.83 $ & $12.0 $   \\ \cline{2-6}
& $2.0$ & $2.0$ & $2.0$ & $1.41$ & $3.0$ \\ \cline{2-6}
& $5.0$ & $5.0$ & $5.0$ & $ 0.89$ & $1.2$ \\ \cline{2-6}
& $9.0$ & $9.0$ & $9.0$ & $ 0.67$ & $ 0.67$ \\ \hline
\end{tabular}
\end{minipage}}
	\caption{Moments of different distributions shown in Figure \ref{fig:dist}.}\label{tab:moments}
\end{table}

\begin{figure}[t]
  \begin{center}
  \includegraphics[width=\textwidth]{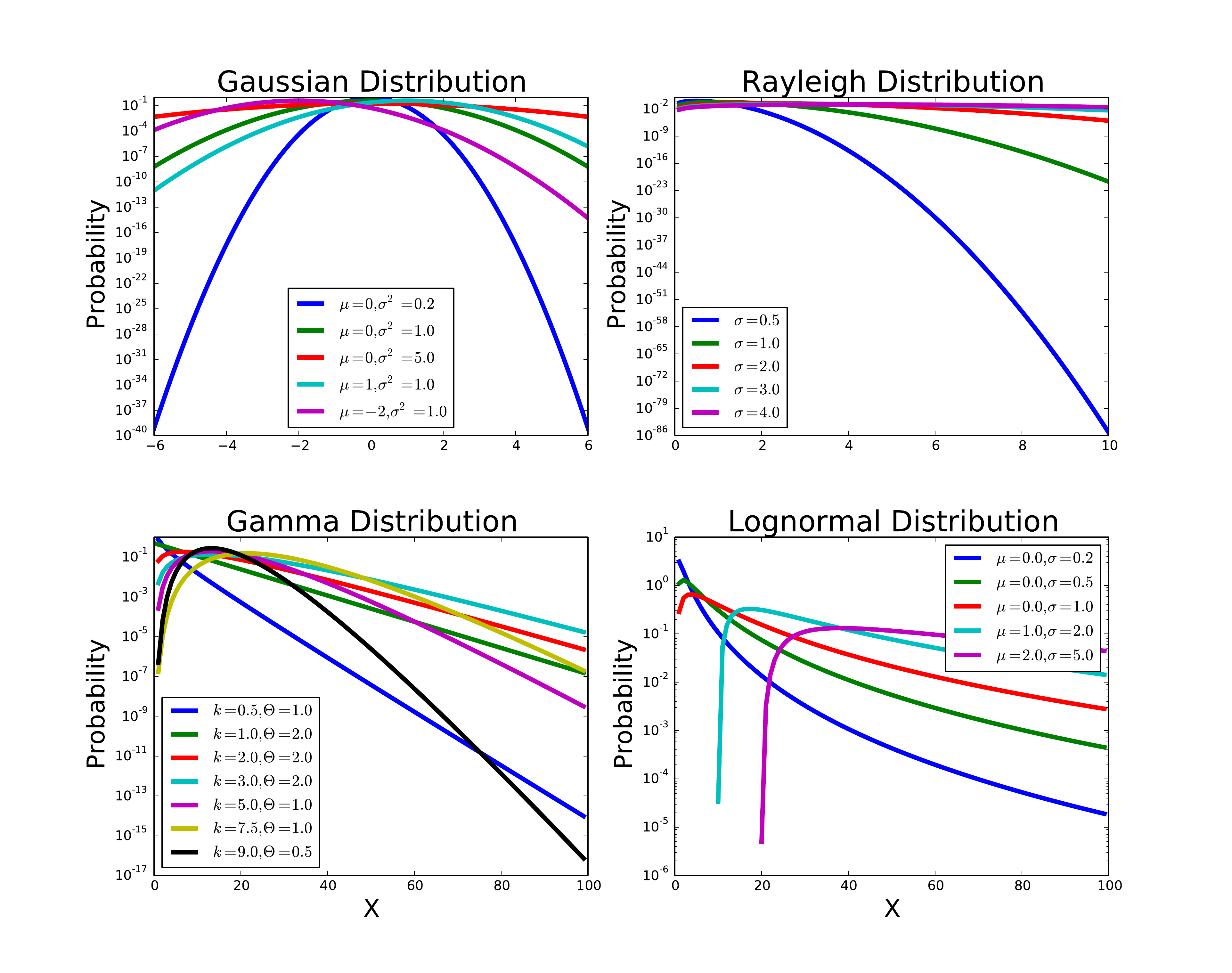}
  \end{center}
  \vspace{-20pt}
  \caption{All distributions explained above (y-axis is in log scale).}\label{fig:dist}
\end{figure}

\subsection{Lognormal Distribution}

 If $x$ is a random variable distributed lognormally then $y = \ln (x)$ is normally distributed where $\ln$ in natural log. It is very helpful in distinguishing the dataset in this work because it gives the best fit so far. The probability density function is given by equation (\ref{eq:lognorm})
\begin{equation}\label{eq:lognorm}
f(x) = \frac{e^{-((\ln((x-\theta)/m))^{2}/(2\sigma^{2}))}} {(x-\theta)\sigma\sqrt{2\pi}} \qquad   x > \theta; m, \sigma > 0
\end{equation} 
where, \\
$\sigma$ = shape parameter (standard deviation)\\
$\theta$ = location parameter\\
$m$ = scale parameter (median)\\

The standard lognormal distribution is given when $\theta = 0 \ \& \  m=1$ and is denoted as equation (\ref{eq:std_lognorm})
\begin{equation}\label{eq:std_lognorm}
f(x) = \frac{e^{-((\ln x)^{2}/2\sigma^{2})}} {x\sigma\sqrt{2\pi}} \qquad x > 0; \sigma > 0
\end{equation}

The lognormal distribution is commonly characterised with its mean $\mu$ given as $\mu = \log m$, using this parameter the density function changes to equation (\ref{eq:lognorm_mu})
\begin{equation}\label{eq:lognorm_mu}
f(x) = \frac{e^{-(\ln(x - \theta) - \mu)^2/(2\sigma^2)}}{(x - \theta)\sigma\sqrt{2\pi}} \qquad x > 0; \sigma > 0
\end{equation}

 Maximum likelihood estimation of this distribution is given by equation (\ref{eq:mle_lognorm}),(\ref{eq:mle_lognorm1}),(\ref{eq:mle_lognorm2})
\begin{align}
\hat{\mu} &= \frac{1}{N} \sum_{i=1}^{N}{\ln{X_i}} \label{eq:mle_lognorm}\\
\hat{\sigma}^2 &= \frac{1}{N} \sum_{i=1}^{N}{(\ln{(X_{i})} - \hat{\mu})^{2}}\label{eq:mle_lognorm1}\\
\hat{m} &= \exp{\hat{\mu}}\label{eq:mle_lognorm2}
\end{align}

Table~\ref{tab:moments} and Figure~\ref{fig:dist} shows the moments of the above mentioned distribution.

\section*{Appendix II}\label{Appendix II}
\begin{figure}[h]
\centering
\includegraphics[width=\textwidth]{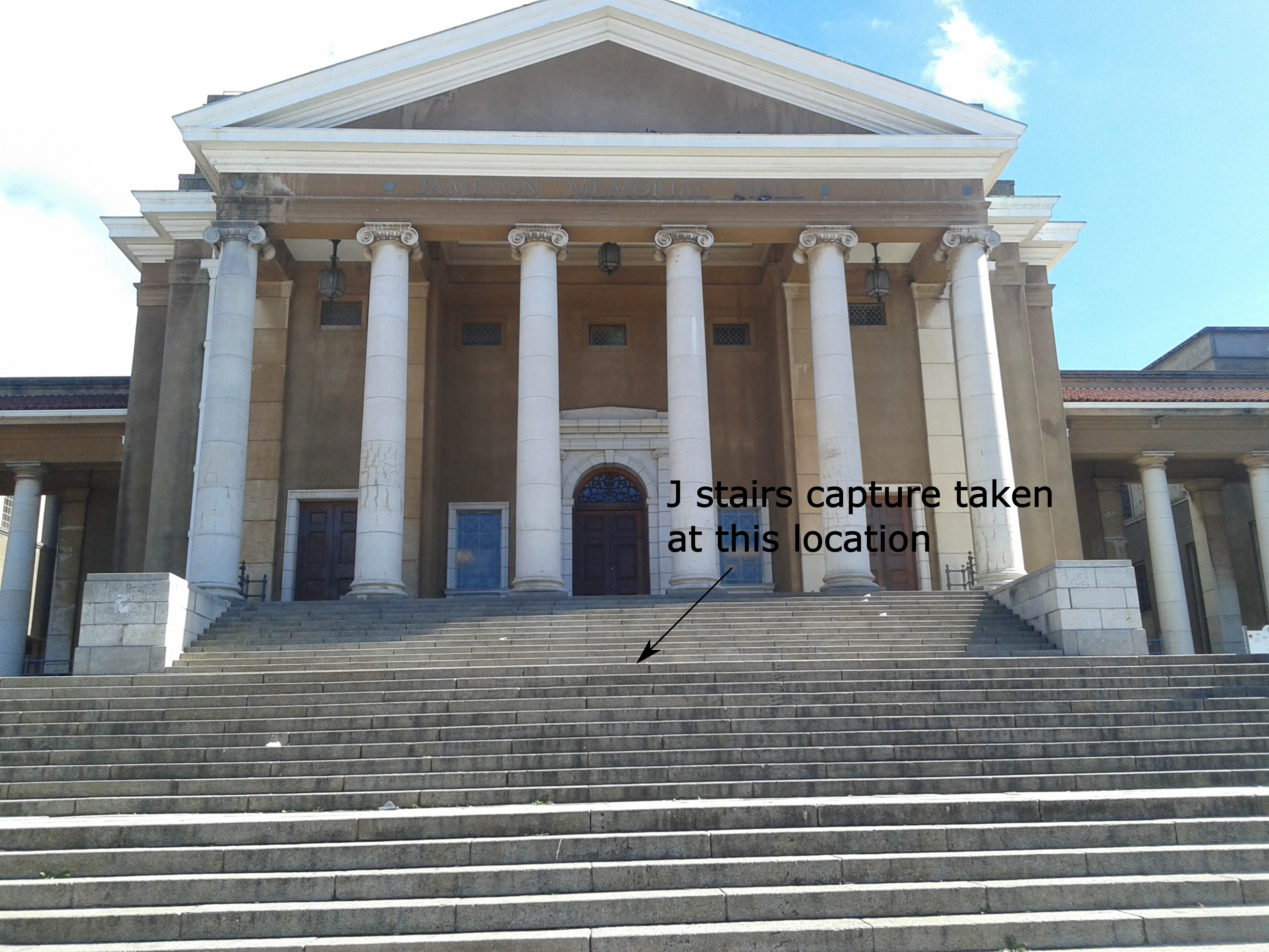} 
	\caption{Location for capture named ``J Stairs'' or ``jameson\_stairs'' or ``jameson\_stairs2''}
		\label{fig:j_stairs}
\end{figure}

The capture named ``J Stairs'' or ``jameson\_stairs'' or ``jameson\_stairs2'' is taken at a location in UCT as shown in Figure~\ref{fig:j_stairs}.

\bibliography{central-bibliography}
\end{document}